# Highly Efficient Blue Host-Free and Host-Guest Organic Light-Emitting Diodes Based on Carbene-Metal-Amides

Patrick J. Conaghan,[1] †, Campbell S. B. Matthews,[1] † Florian Chotard,[2] Saul T. E. Jones[1], Neil C. Greenham,[1] Manfred Bochmann,[2] Dan Credgington[1]* and Alexander S. Romanov[2]*

[1] P. J. Conaghan, C. S. B. Matthews, Prof. Dr. N. Greenham, Dr. D. Credgington, Cavendish Laboratory, Department of Physics, University of Cambridge, J J Thomson Avenue, CB3 0HE, UK

[2] Dr. A. S. Romanov, Dr. F. Chotard, Prof. Dr. M. Bochmann, School of Chemistry, University of East Anglia, Earlham Road, Norwich, NR4 7TJ, UK

* Correspondence to djnc3@cam.ac.uk, A.Romanov@uea.ac.uk

† These authors contributed equally

**Abstract:** Carbene-metal-amide type photoemitters based on $CF_3$-substituted carbazolate ligands show sky-blue to deep-blue photoluminescence from charge-transfer excited states. They are suitable for incorporation into organic light-emitting diodes (OLEDs) by thermal vapour deposition techniques, either embedded within a high-triplet-energy host, or used host-free. We report high-efficiency OLEDs with emission ranging from yellow to blue (Commission Internationale de l'Éclairage (CIE) coordinates from [0.35, 0.53] to [0.17, 0.17]). The latter show a peak electroluminescence external quantum efficiency (EQE) of 20.9 % in a polar host. We observe that the relative energies of CT and $^3$LE states influence the performance of deep-blue emission from carbene-metal-amide materials. We report prototype host-free blue devices with peak external quantum efficiency of 17.3 %, which maintain high performance at brightness levels of 100 cd m$^{-2}$.

**Main Text:** The development of organic light-emitting diodes (OLEDs) has continued for over 30 years since the discovery of electroluminescence from the fluorescent green emitter tris-(8-hydroxyquinoline)aluminium.[1] An OLED emits light via radiative recombination of strongly-bound excitons formed from electrically-injected charge pairs. For injected charges with uncorrelated spins, excitons are generated in a 3:1 ratio of spin-1 triplet (T) and spin-0 singlet (S) states. In the absence of spin-orbit coupling, only the singlets can relax to the ground state via photon emission, and the energy of triplets is wasted.[2] Luminescent harvesting of triplet states can be achieved in several ways. Use of heavy metal complexes, notably of iridium(III) and platinum(II) with high spin-orbit coupling enables photoemission via phosphorescence on microsecond time scales.[3–7] Alternatively, compounds can be designed where the lowest energy excited states, $S_1$ and $T_1$, are close in energy, such that thermal equilibrium between the two at typical operating temperatures enables emission from the $S_1$ state via (reverse) intersystem crossing. This process is termed E-type (or Thermally Activated) Delayed Fluorescence.[8–10] The development of materials for blue OLEDs however remains particularly challenging, with high-energy bimolecular interactions involving long-lived triplet excitons implicated as one of the primary limits to operational lifetime.[11]

We have shown recently that linear, two-coordinated coinage metal complexes of the type (L)MX (M = Cu, Ag, Au) can show efficient photoluminescence (PL) via a delayed fluorescence mechanism, provided the ligand L is a strongly-bound carbene capable of acting as π-acceptor and X is an electron-rich anion capable of acting as electron donor upon excitation.[12–16] Combinations of L = cyclic (alkyl)(amino)carbene (CAAC) [17,18] and X = carbazolate (Cz) have proved particularly successful and have become known as "carbene-metal-amide" (CMA)-type photoemitters.[19] These complexes are soluble in most organic solvents and sufficiently thermally stable to enable the fabrication of OLED devices by both solution processing and thermal vapour deposition techniques.[13,19–21] CMAs do not suffer from strong concentration quenching in the solid state, attributed to the lack of close metal-metal contacts, enabling the realization of host-free green devices with an external quantum efficiency (EQE) of 23.1%.[20] Due to the linear two-coordinate geometry, CMAs are conformationally flexible, with a low barrier for rotation about the metal-N σ-bond. The highest occupied molecular orbital (HOMO) is centred on the amide ligand while the lowest unoccupied molecular orbital (LUMO) consists mainly of the p-orbital of the carbene C atom, with only a small metal contribution to both. Excitation of the molecule thus leads to a charge-transfer (CT) type excited state. Both frontier orbitals are spatially well-separated, commensurate with a small $\Delta E(S_1-T_1)$ energy gap for CT excitations. This arrangement enables luminescence quantum efficiencies approaching 100% coupled with short, sub-microsecond excitation lifetimes for triplet states. The emission process involved has been subject to a number of theoretical and spectroscopic investigations;[22–24] modelling has shown that upon rotation about the metal-N σ-bond, the $\Delta E(S_1-T_1)$ energy gap decreases and at high twist angles may approach zero.[13,21–24] Here we report the synthesis of new Au-bridged emitters which enable blue host-free OLEDs with EQEs of 17.3% ($\lambda_{em,max}$ = 473 nm), as well as host:guest devices with a peak wavelength of 450 nm and an EQE of up to 20.9%. At practical brightness levels of 100 cd $m^{-2}$, we achieve $\eta_{EQE}$ = 17.2% and $\eta_{EQE}$ = 17.8 % for the best host-free and host:guest devices, respectively.

As a consequence of the high polarity of CMA emitters, emission energies are sensitive to their molecular environment; for example, this has allowed "tuning" of electroluminescence by suitable host media from green to sky-blue.[20] However, much larger changes in emission energies can be achieved by altering the carbazole substitution pattern. For the ($^{Ad}$CAAC)AuCz archetype, the HOMO is almost entirely located on the Cz donor. The introduction of electron-withdrawing groups to this moiety therefore influences the HOMO more than the LUMO and widens the HOMO-LUMO gap, resulting in a shift towards blue emission. We therefore prepared the $CF_3$-substituted carbazolate complexes **1** ($R^1 = CF_3$, $R^2 = {}^tBu$) and **2** ($R^1 = R^2 = CF_3$) (Figure 1) from ($^{Ad}$CAAC)AuCl and the corresponding carbazoles in the presence of KO$^t$Bu, following previously published procedures.[19–21] The structural and PL data of the known green ($R^1 = R^2 = H$, here complex **3**) and yellow emitters ($R^1 = R^2 = {}^tBu$, here complex **4**) are included for comparison.[19] Complexes **1** and **2** were prepared on a 5 g scale as white solids which are stable in air and in solution for long periods of time. They possess good solubility in aromatic solvents (toluene, chlorobenzene, 1,2-difluorobenzene), THF, dichloromethane, or DMF but are insoluble in hexane. We chose the adamantyl-substituted CAAC ligand ($^{Ad}$CAAC) since its high steric hindrance provides good thermal stability and high PL intensities. Thermogravimetric analysis (TGA, 5% weight loss) showed decomposition temperatures of 325 °C for **1** and 364 °C for **2** (Fig. S1).

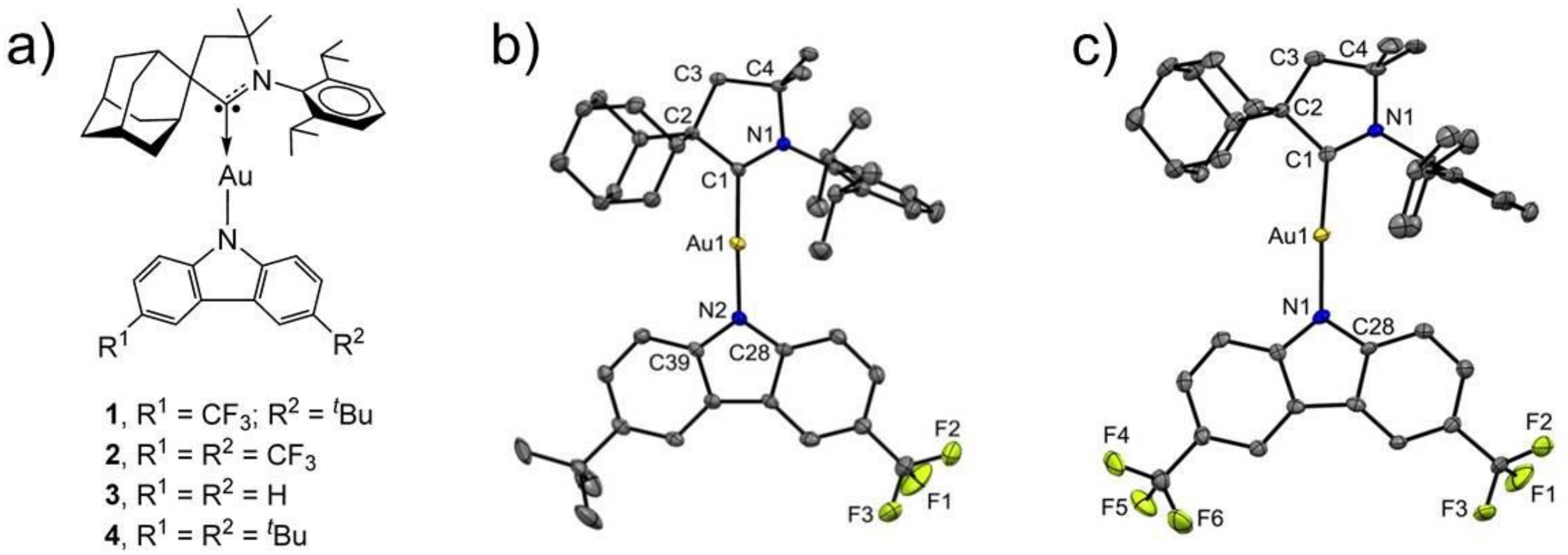

**Figure 1.** (a) Structures of carbene metal amides **1**, **2**, **3** and **4** and single crystal X-ray structures of **1** (b) and **2** (c). Ellipsoids are shown at 50 % probability.

The structures of **1** and **2** were confirmed by single crystal X-ray diffraction (Figure 1). Both complexes show a two-coordinate linear geometry for the gold atom with negligible deviation from linearity (Table 1); there are no close Au···Au contacts. The HOMO-LUMO overlap is directly related to the donor-acceptor distance C1(CAAC)···N2(Cz), which is 0.02 Å shorter in **1** than in **2**. This is likely to impact radiative rates and $\Delta E(S_1\text{-}T_1)$.[13,25] The torsion angle N1–C1–N2–C28 for **1** of 14.3° is similar to those in **3**/**4**, whereas for **2** it is nearly 0°, possibly due to intermolecular interactions in the crystal (Figure S2).

**Table 1.** Selected bond lengths [Å] and angles [°] of **1** and **2**. Measurements of **3** and **4** are reproduced for comparison (average values for the two independent molecules in the unit cell for **1** and **2**).

| | Au–C1 | Au–N2 | C1···N2 | angle C1–Au–N2 | torsion angle N1–C1–N2–C28 |
|---|---|---|---|---|---|
| **1** | 1.983(4) | 2.016(3) | 3.999(4) | 177.54(15) | 14.3(4) |
| **2** | 1.994(5) | 2.024(4) | 4.018(4) | 176.06(16) | 1.1(2) |
| **3** | 1.991(3) | 2.026(2) | 4.017(3) | 178.78(11) | 17.7(1) |
| **4** | 1.997(3) | 2.020(2) | 4.017(3) | 178.25(11) | 16.3(1) |

The redox behaviour of **1** and **2** was analysed in acetonitrile solution using $[^{n}Bu_4N]PF_6$ as the supporting electrolyte (see SI for cyclic voltammograms, Figure S3). The electrochemical data are collected in Table 2. Both **1** and **2** show a quasi-reversible, carbene ligand-centred reduction processes. Complex **1** has the smallest peak-to-peak separation ($\Delta E_p$) in the series (73 mV), indicating higher stability of the reduced species. The reduction potential is sensitive to the number of Cz-$CF_3$ groups and leads to a greater LUMO stabilization for **1** and **2** compared with **3/4**. Both **1** and **2** show irreversible carbazole-centered oxidation processes (Figure S3). The HOMO levels from the onset of the first oxidation potentials[26] are –5.85 and –6.12 eV for **1** and **2**, respectively, compared to –5.61 eV for **3** and –5.47 eV for **4**.

**Table 2.** Formal electrode potentials (peak position $E_p$ for irreversible and $E_{1/2}$ for quasi-reversible processes (*), *V*, *vs*. $FeCp_2$), onset potentials (*E*, *V*, *vs*. $FeCp_2$), peak-to-peak separation in parentheses for quasi-reversible processes ($\Delta E_p$ in mV), $E_{HOMO}/E_{LUMO}$ (*e*V) and band gap values ($\Delta E$, *e*V) for the investigated complexes.[a]

| Complex | Reduction | | $E_{LUMO}$ eV | Oxidation | | | $E_{HOMO}$ eV | $\Delta E$ eV |
|---|---|---|---|---|---|---|---|---|
| | $E_{1st}$ | $E_{onset\ red}$ | | $E_{1st}$ | $E_{onset\ ox}$ | $E_{2nd}$ | | |
| **1** | –2.65* (73) | –2.57 | –2.82 | +0.57 | +0.46 | +1.05 | –5.85 | 3.03 |
| **2** | –2.55* (90) | –2.47 | –2.92 | +0.83 | +0.73 | – | –6.12 | 3.20 |
| **3** | –2.68* (80) | –2.60 | –2.79 | +0.26 | +0.22 | +0.77 | –5.61 | 2.82 |
| **4** | –2.86* (83) | –2.78 | –2.61 | +0.13 | +0.08 | +0.65 | –5.47 | 2.86 |

[a] Measured in THF solution, recorded using a glassy carbon electrode, concentration 1.4 mM, supporting electrolyte $[^{n}Bu_4N][PF_6]$ (0.13 M), measured at 0.1 V $s^{-1}$; $E_{HOMO} = -(E_{onset\ ox\ Fc/Fc+} + 5.39)$ *e*V; $E_{LUMO} = -(E_{onset\ red\ Fc/Fc+} + 5.39)$ eV. Measurements of **3** and **4** are reproduced for comparison.

The UV/vis absorption spectra of **1** and **2** were measured in THF (Figure 2). All complexes show π–π* transitions at ca. 270 nm which can be ascribed to intra-ligand (IL) transitions of the CAAC carbene, and weaker progressions around 300-310 nm and 360-375 nm ascribed to π–π* IL transitions of the amide. The broad, low-energy absorption band is assigned largely to ligand-to-ligand CT transitions {π(carbazole)–π*(CAAC)}. The onset of the absorption CT band and its peak position exhibit negative solvatochromism, blue-shifting by *ca.* 20 nm on each step of the series **4→3→1→2** with decreasing electron-donor character of the amide ligand. The observed trend in absorption spectra is largely consistent with the increase of the band gap ($\Delta E$, Table 2) identified by cyclic voltammetry (see below).

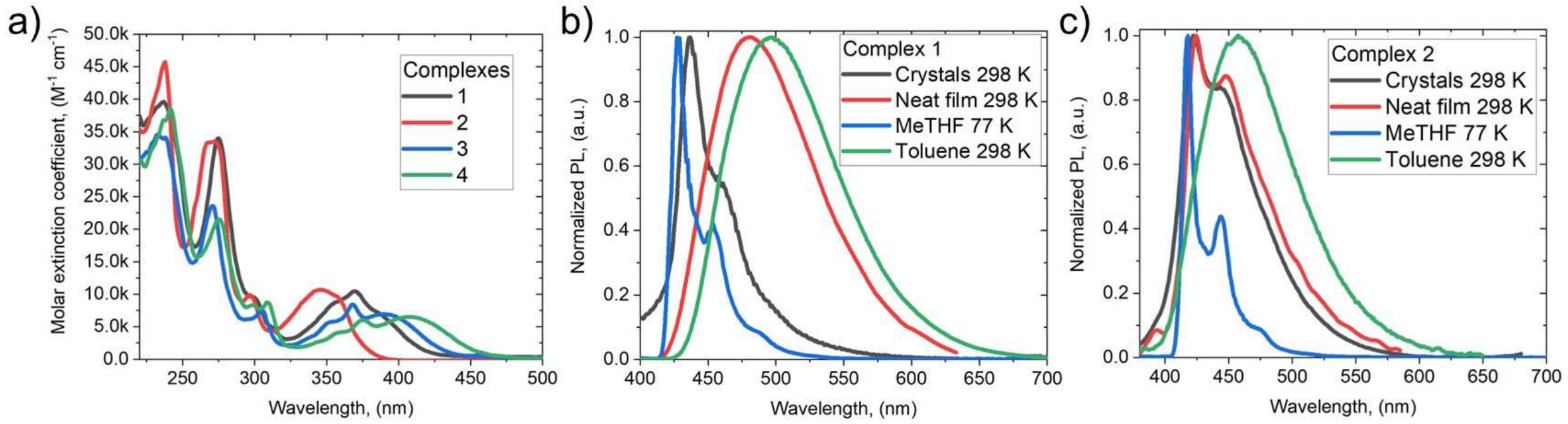

**Figure 2.** (a) UV-vis spectra in THF solution for gold complexes **1**, **2** in comparison with **3** and **4**. PL spectra of (b) **1** and (c) **2** at 298K as crystals, in neat film, in frozen MeTHF solution and in liquid toluene solution (excitation at 365 nm).

On excitation with UV light ($\lambda_{exc}$ = 365 nm), **1** and **2** as crystals show blue photoluminescence at $\lambda_{max}$<440 nm, dominated by structured emission. This emission is similar to that observed for all complexes in frozen 2-MeTHF at 77 K (Figure S4). We ascribe the structured emission to local (i.e. ligand-centred) triplet excited states ($^3$LE). The behaviour of **1** and **2** contrasts with that of **3** and **4** (Figure S5), which exhibit unstructured CT emission in the crystalline phase. Emission red-shifts in neat amorphous thin films and in liquid solution, becoming broad and unstructured. We ascribe this emission to luminescence from CT excited states, which represent the lowest-energy triplet excitation in these less-constrained environments. The energy of the CT transition ($E_{CT}$), measured from its high-energy onset in toluene solution, increases with increasing electron-acceptor strength of the carbazole substituents, from 2.62 eV for **4** to 3.15 eV for **2**. The effect of carbazole substitution on the $^3$LE emission energy ($E_{LE}$), measured from its high-energy onset in frozen MeTHF solution, is less pronounced, with a change of only 0.11 eV between **4** and **2** (Table 3).

The CT excited state energy and lifetime of **1** are dependent on its environment (Table 3). At 298 K in toluene solution, the emission lifetime is 740 ns. Peak emission energy blue-shifts by 60 meV in neat solid samples and the lifetime increases to 1 μs. In DPEPO host, which increases peak emission energy by a further 110 meV, the lifetime increases to around 20 μs. On cooling to 77K, lifetimes of solid samples increase to 600-700 μs.

For **2**, excited state lifetimes at 298 K follow a similar trend and are longer. Emission in toluene solution has a lifetime of 11.5 μs and exhibits CT character. In neat solid film, emission blue-shifts, structure associated with $^3$LE emission is observed and relative luminescence yield decreases markedly. On cooling solid samples to 77 K, lifetimes increase to over 2 ms and the $^3$LE character of emission becomes dominant.

The PL quantum yield in solution (298 K) drops from near unity for **1** to 61% for **2**. PL is reversibly suppressed in the presence of $O_2$, as expected for emission via triplet states. We thus observe that emission lifetime is correlated with $E_{CT}$, increasing as the $\Delta E$(CT-$^3$LE) gap narrows. $^3$LE phosphorescence is observed in environments where the lowest CT states are no longer the lowest-energy triplet excitations, and PL quantum yield is reduced.

OLEDs utilising complexes **1-4** as emitters were fabricated by thermal vapour deposition under high vacuum ($10^{-7}$ Torr) on ITO-coated glass substrates with sheet resistance of 15 Ω/□. Two device architectures were employed, shown in Figure 3 together with chemical structures for the materials used. Architecture A was used for the blue-emitting complexes **1** and **2**. A 40 nm layer of 1,1-bis{4-[*N*,*N*-di(4-tolyl)amino]phenyl}cyclohexane (TAPC) functions as a hole transport layer, with a 5 nm layer of 9,9′-biphenyl-2,2′-diylbis-9H-carbazole (o-CBP) acting as an exciton blocking layer due to its slightly higher $T_1$ energy and deeper highest occupied molecular orbital (HOMO). The 30 nm thick emissive layer (EML) was composed of either pure **1** and **2** in a host-free configuration or with the emitting material doped at 20 weight-% into a bis[2-diphenyl-phosphino)phenyl]ether oxide (DPEPO) host. A 40 nm layer of diphenyl-4-triphenylsilyl-phenylphosphine oxide (TSPO1) was used as the electron-transporting and hole-blocking layer.

Devices containing the yellow emitter **4** and green emitter **3** in a host-free configuration or doped at 20 wt.% in either DPEPO or 1,3,5-tris(carbazol-9-yl)benzene (TCP) were fabricated according to architecture B using a 10 nm layer of 1,4-bis(triphenylsilyl)benzene (UGH2) as a hole-blocking layer and 1,3,5-tris(2-N-phenylbenzimidazole-1-yl)benzene (TPBi) as electron-transport layer.

**Table 3.** Emission data of **1, 2, 3** and **4** in toluene solution and in thermally evaporated solid films.

| Complex | **1** | | | | **2** | | **3** | | | | **4** | | | |
|---|---|---|---|---|---|---|---|---|---|---|---|---|---|---|
| | toluene | neat | o-CBP:**1** | DPEPO:**1** | toluene | neat | toluene | neat[b] | TCP:**3**[b] | DPEPO:**3** | toluene | neat | TCP:**4** | DPEPO:**4** |
| $\lambda_{em}$ (nm) | 495 | 484 | 479 | 464 | 456 | 425 | 528 | 500 | 500 | 485 | 552 | 540 | 535 | 514 |
| $\tau$ (µs) | 0.74 | 0.99 | 1.05 | 19.4 | 11.5 | 10.8 | 1.25 | 0.76 | 0.97 | 1.31 | 0.84 | 0.69 | 0.98 | 1.12 |
| $\Phi$(%, 300K; $N_2$) | 96 | - | - | - | 61 | - | 98 | - | - | - | 95 | - | - | - |
| $k_r$ ($10^5$ $s^{-1}$) | 13.1 | - | - | - | 0.53 | - | 7.8 | - | - | - | 11.3 | - | - | - |
| $k_{nr}$ ($10^5$ $s^{-1}$) | 0.54 | - | - | - | 0.34 | - | 0.16 | - | - | - | 0.60 | - | - | - |
| $^1$CT/$^3$LE (eV)[a] | 2.86/2.97 | | | | 3.15/3.03 | | 2.76/2.96 | | | | 2.62/2.92 | | | |
| $\Delta E(^1CT–^3LE)$(eV)[a] | –0.11 | | | | 0.12 | | –0.20 | | | | –0.30 | | | |

[a] $^1$CT and $^3$LE energy levels based on the onset values of the emission spectra blue edge at 77 K in MeTHF glass and in solution at 298 K. [b] reproduced from [20]

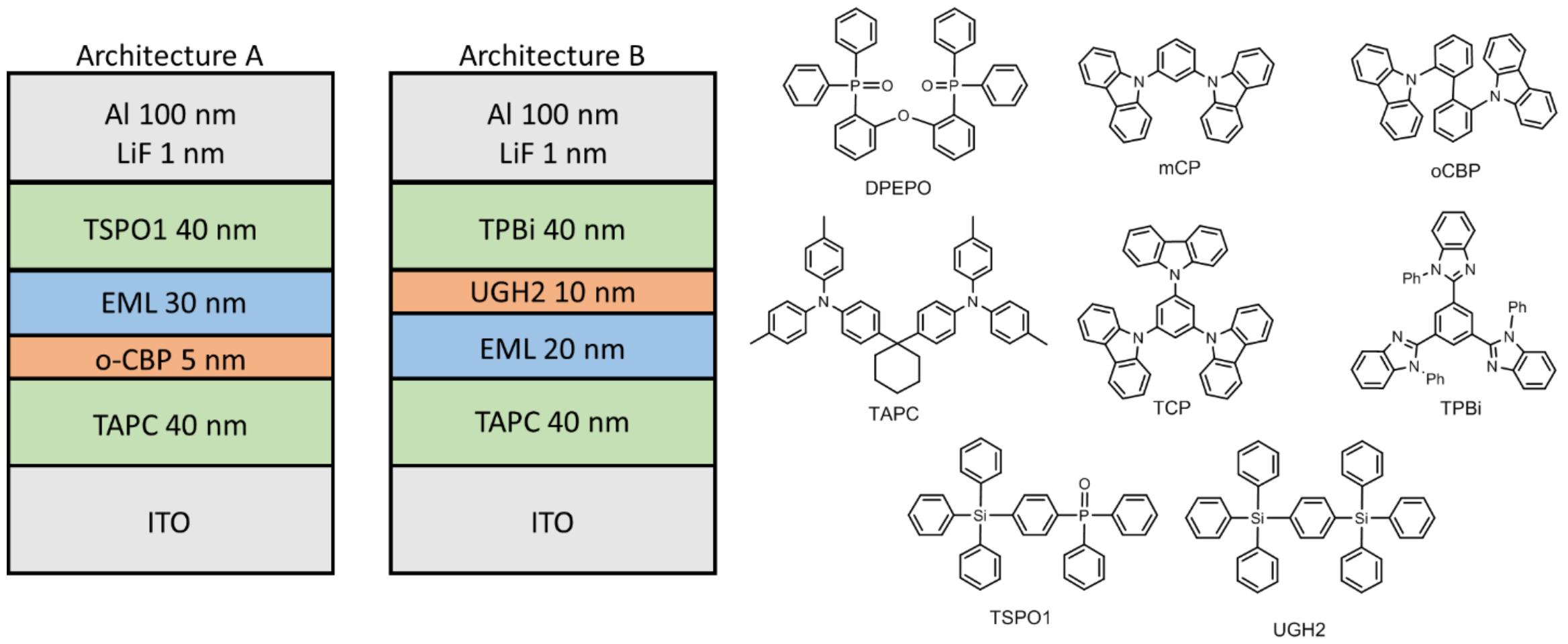

**Figure 3.** OLED architectures A (for **1** and **2**) and B (for **3** and **4**) and chemical structures of the materials used.

Electroluminescence spectra for devices based on **1** are shown in Figure 4; peak wavelengths ($\lambda_{Peak}$) and Commission Internationale de l'Éclairage (CIE) colour space co-ordinates are summarized in Table 4. Host-free **1** devices show $\lambda_{Peak}$ = 473 nm and CIE (0.18, 0.27). The electroluminescence can be blue-shifted by dispersal in a low-polarity host material, here o-CBP, and further shifted by utilising a high-polarity host material, here DPEPO, to obtain a peak wavelength of 450 nm (CIE 0.17, 0.17).

Similar shifts are achievable for **3** and **4**, utilising TCP / DPEPO as the low / high polarity host material, respectively, in agreement with our previous report of **3** in mCP host. Host-free **2** devices show deep-blue electroluminescence with $\lambda_{Peak}$ = 423 nm and CIE (0.16, 0.05); however, emission appears to be of $^3$LE character (Figure S6) and the devices undergo rapid degradation, with an additional spectral feature at 580 nm arising on a ~1 s timescale. Full device characterisation was therefore only carried out using **1, 3** and **4** as emitters.

Devices based on **1** show low turn-on voltages (reported as the applied bias at which luminance equals 1 cd m$^{-2}$) of $V_{On}$ = 3.7 V for **1**, which indicates an absence of large barriers to charge injection into the emitting layer even in host-free architectures. Figure 5 shows the current density-voltage and luminance-voltage characteristics of champion **1** devices in both host-free and host:guest environments. The external quantum efficiency ($\eta_{EQE}$) values as a function of current density (Figure 6) were calculated from on-axis irradiance measurements at a small solid angle assuming Lambertian emission, as is commonly observed for planar OLEDs.[20] The peak $\eta_{EQE}$ = 17.3 % for blue host-free **1** devices is an indication of the insensitivity of gold-bridged

CMA materials to aggregation-quenching. The efficiency rises slightly to $\eta_{EQE}$ = 20.9 % on dilution in a DPEPO host. Host free devices exhibit reduced roll-off compared to host-guest devices, as might be expected from a reduction in triplet density under operation. Equivalent data for compounds 3 and 4 are shown in figures S6-8.

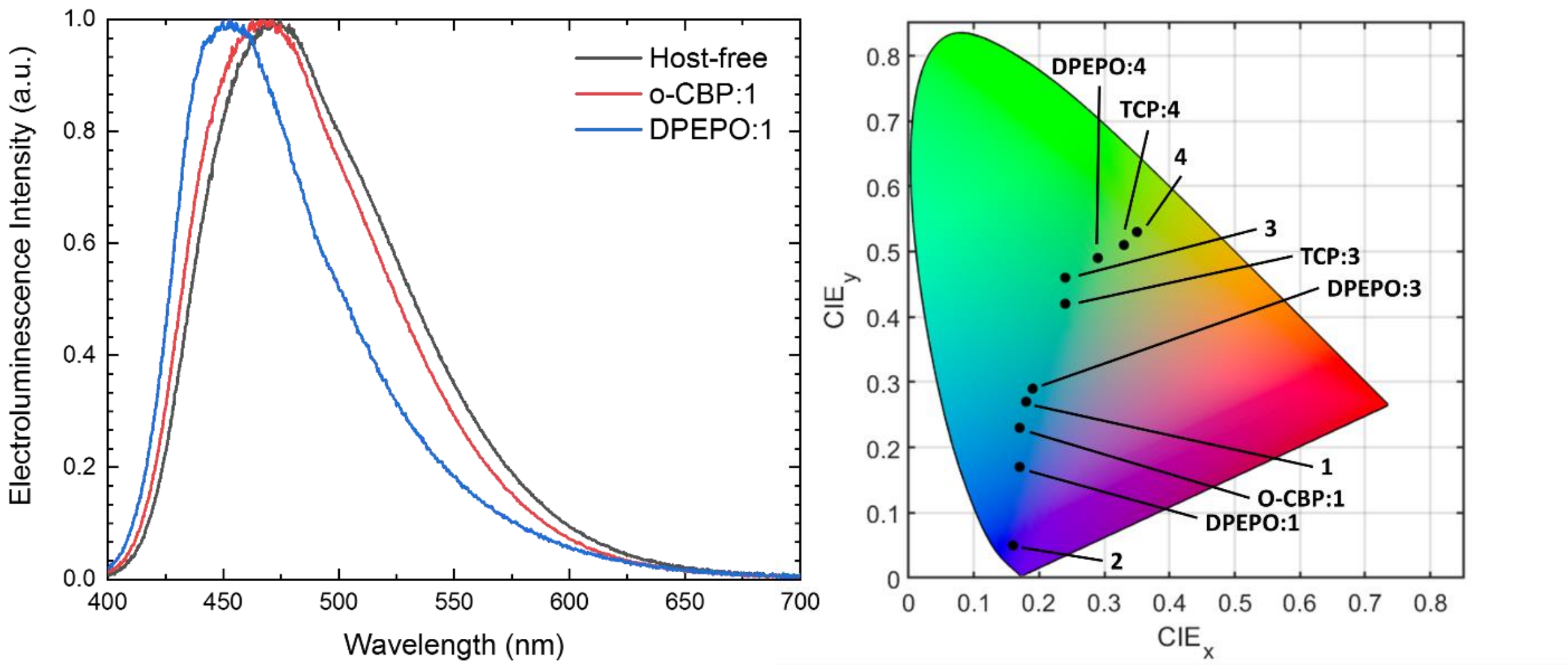

**Figure 4** Left: Normalised electroluminescence spectra from devices incorporating **1** in host-free and host-guest structures. Right: CIE colour space chart depicting the apparent colour of electroluminescence from OLED devices incorporating CMA emitters in host-free and host:guest structures. CIE coordinates for **3** in TCP host and host-free are reproduced from ref. [20].

**Table 4.** Summary of champion OLED turn-on voltage, $\eta_{EQE}$ performance and spectral parameters for varying EML composition. Metrics for **3** and TCP:**3** OLEDs reproduced from ref. [20]

| Emitting Layer | $V_{On}$ [V]$^a$ | $\eta_{EQE}$ [%] (max.) | $\eta_{EQE}$ [%] (100 cd m$^{-2}$) | $\lambda_{Peak}$ [nm] | CIE (x.y) |
|---|---|---|---|---|---|
| **1** | 3.7 | 17.3 | 17.2 | 473 | (0.18, 0.27) |
| o-CBP:**1** | 3.9 | 17.2 | 16.0 | 466 | (0.17, 0.23) |
| DPEPO:**1** | 3.7 | 20.9 | 17.8 | 450 | (0.17, 0.17) |
| **2** | - | - | - | 423 | (0.16, 0.05) |
| **3** | 3.5 | 23.1 | 23.0 | 500 | (0.24, 0.46) |
| TCP:**3** | 3.3 | 26.9 | 26.4 | 500 | (0.24,0.42) |
| DPEPO:**3** | 4.4 | 21.9 | 20.8 | 474 | (0.19, 0.29) |
| **4** | 5.1 | 11.5 | 10.1 | 536 | (0.35, 0.53) |
| TCP: **4** | 4.7 | 18.7 | 18.4 | 527 | (0.33, 0.51) |
| DPEPO: **4** | 4.7 | 24.7 | 22.9 | 518 | (0.29, 0.49) |

$^a$ Values at brightness >1 cd/m$^2$.

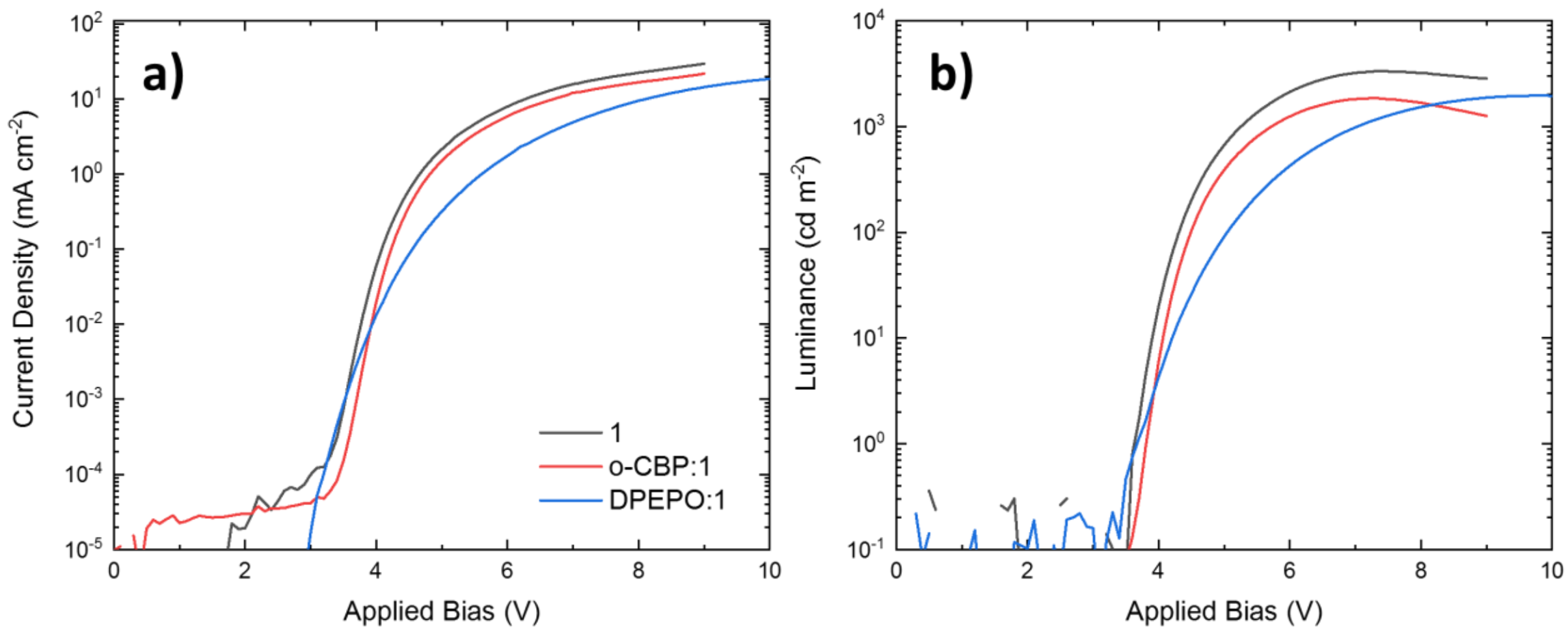

**Figure 5** a) Current density-voltage and b) luminance-voltage characteristics for OLEDs based on **1** in host-free and host:guest environments.

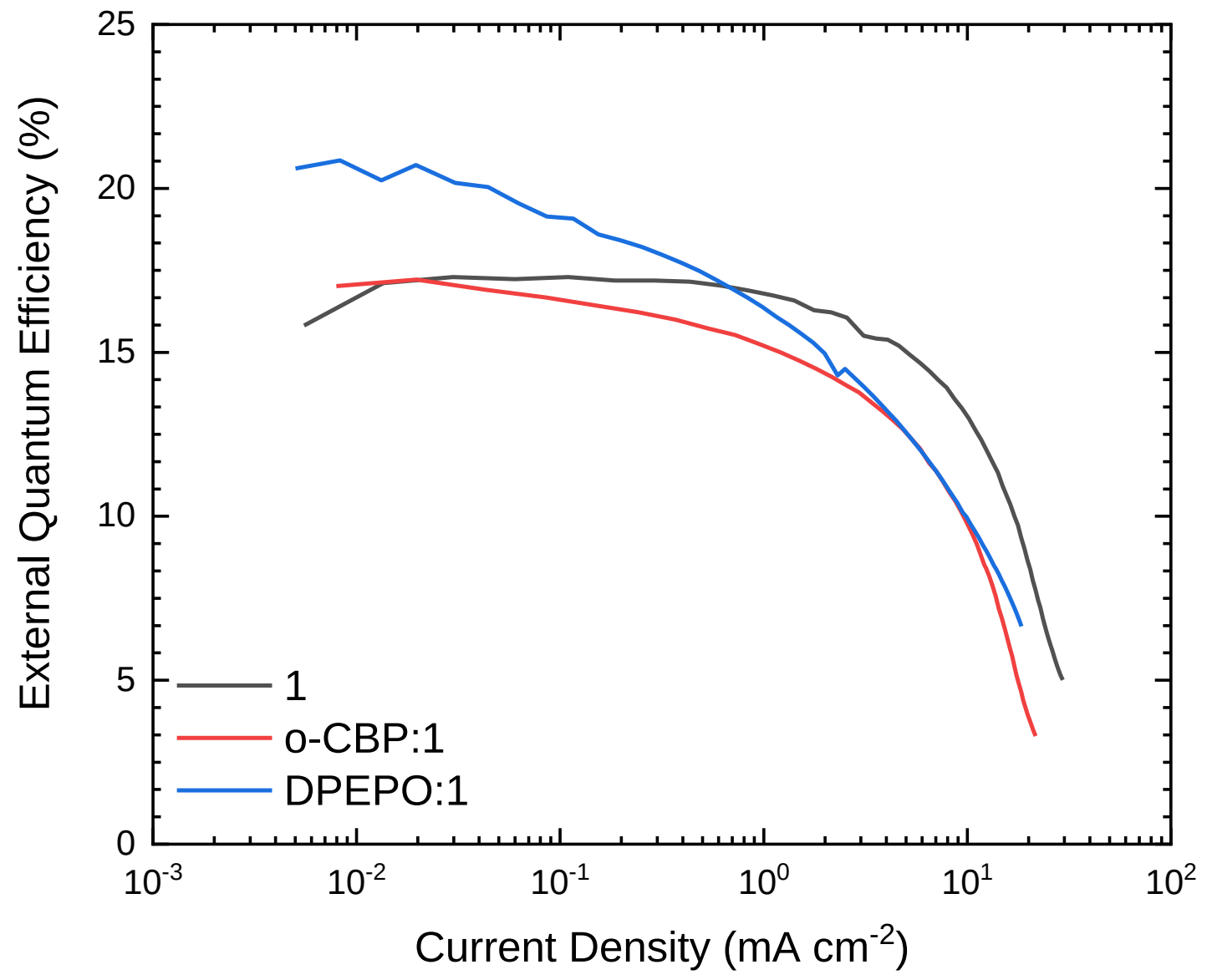

**Figure 6** External quantum efficiency of OLEDs based on **1** as a function of current density in host-free and host:guest environments

In summary, by varying the electron donating or withdrawing nature of carbazole substituents the HOMO-LUMO gap of CMA-type photoemitters can readily be adjusted and the electroluminescence colour can be tuned over a wide range, from yellow (CIE co-ordinates 0.35, 0.53) to deep-blue (CIE 0.16, 0.05). In addition, the CT energy is sensitive to the host environment. These effects enable control of the energy gap between CT and LE excitations. From transient photoluminescence measurements we establish that rapid triplet luminescence is correlated with a larger $\Delta E$(CT-$^3$LE) gap. Approaching resonance between the CT state and the amide triplet at around 3 eV leads to a marked reduction in performance, in contrast with current models of all-organic TADF luminescence. The effect is most clearly seen in the case of complex **2** with the most electronically depleted carbazole ligand, where in some environments the $^3$LE is the lowest-lying triplet. Where this is the case, long-lived structured phosphorescence is observed and device performance is reduced. However, the combined approach of tuning molecular design and host environment allows devices to operate below this ceiling. We have fabricated efficient blue host-guest OLEDs with peak EQE of 20.9 % and peak electroluminescence wavelength of 450 nm (CIE co-ordinates [0.17, 0.17]). We also achieve peak EQE of 17.3 % for blue OLEDs (CIE [0.18, 0.27]) in host-free architectures. Host-guest and host-free devices exhibit slow roll-off, achieving EQEs at a brightness of 100 cd $m^{-2}$ of 17.8 % and 17.2 %, respectively. The host-free approach can simplify the design of blue OLEDs and may avoid the requirement for stable, wide-gap and high-triplet-energy host materials.

**Acknowledgements**

This work was supported by the Engineering and Physical Sciences Research Council (EPSRC, grant no. EP/M005143/1), the Royal Society, the European Research Council (ERC) and Samsung Display Co. Ltd (SDC). P.J.C acknowledges EPSRC for Ph.D. studentship funding. C.S.B.M acknowledges the Royal Society (RGF\EA\180041) and St John's College, Cambridge. M. B. is an ERC Advanced Investigator Award holder (Grant No. 338944-GOCAT). D.C. and S.T.E.J acknowledge support from the Royal Society (grant nos. UF130278 and RG140472). A.S.R. acknowledges support from the Royal Society (grant nos. URF\R1\180288 and RGF\EA\181008). We thank the National Mass Spectrometry Facility at Swansea University. The data underlying this publication are available through the following web link: [To be inserted at proof]

## References

1. Tang, C. W. & Vanslyke, S. A. Organic electroluminescent diodes. *Appl. Phys. Lett.* **51**, 913–915 (1987).
2. Yersin, H., Rausch, A. F., Czerwieniec, R., Hofbeck, T. & Fischer, T. The triplet state of organo-transition metal compounds. Triplet harvesting and singlet harvesting for efficient OLEDs. *Coord. Chem. Rev.* **255**, 2622–2652 (2011).
3. Baldo, M. A. *et al.* Highly efficient phosphorescent emission from organic electroluminescent devices. *Nature* **395**, 151–154 (1998).
4. Baldo, M. A., Adachi, C. & Forrest, S. R. Transient analysis of organic electrophosphorescence. II. Transient analysis of triplet-triplet annihilation. *Phys. Rev. B* **62**, 10967–10977 (2000).
5. Evans, R. C., Douglas, P. & Winscom, C. J. Coordination complexes exhibiting room-temperature phosphorescence: Evaluation of their suitability as triplet emitters in organic light emitting diodes. *Coord. Chem. Rev.* **250**, 2093–2126 (2006).
6. Schulz, L. *et al.* Importance of intramolecular electron spin relaxation in small molecule semiconductors. *Phys. Rev. B - Condens. Matter Mater. Phys.* **84**, 085209 (2011).
7. Li, T. Y. *et al.* Rational design of phosphorescent iridium(III) complexes for emission color tunability and their applications in OLEDs. *Coordination Chemistry Reviews* (2018). doi:10.1016/j.ccr.2018.06.014
8. Wong, M. Y. & Zysman-Colman, E. Purely Organic Thermally Activated Delayed Fluorescence Materials for Organic Light-Emitting Diodes. *Advanced Materials* **29**, 1605444 (2017).
9. Bizzarri, C., Hundemer, F., Busch, J. & Bräse, S. Triplet emitters versus TADF emitters in OLEDs: A comparative study. *Polyhedron* **140**, 51–66 (2018).
10. Bizzarri, C., Spuling, E., Knoll, D. M., Volz, D. & Bräse, S. Sustainable metal complexes for organic light-emitting diodes (OLEDs). *Coord. Chem. Rev.* **373**, 49–82 (2018).
11. Giebink, N. C. *et al.* Intrinsic luminance loss in phosphorescent small-molecule organic light emitting devices due to bimolecular annihilation reactions. *J. Appl. Phys.* **103**, 044509 (2008).
12. Romanov, A. S. *et al.* Copper and Gold Cyclic (Alkyl)(amino)carbene Complexes with Sub-Microsecond Photoemissions: Structure and Substituent Effects on Redox and Luminescent Properties. *Chem. - A Eur. J.* **23**, 4625–4637 (2017).

13. Romanov, A. S. *et al.* Mononuclear Silver Complexes for Efficient Solution and Vacuum-Processed OLEDs. *Adv. Opt. Mater.* 1801347 (2018). doi:10.1002/adom.201801347

14. Di, D. *et al.* High-performance light-emitting diodes based on carbene-metal-amides. *Science* **356**, 159–163 (2017).

15. Shi, S. *et al.* Highly Efficient Photo- and Electroluminescence from Two-Coordinate Cu(I) Complexes Featuring Nonconventional N-Heterocyclic Carbenes. *J. Am. Chem. Soc.* **141**, 3576–3588 (2019).

16. Hamze, R. *et al.* Eliminating nonradiative decay in Cu(I) emitters: >99% quantum efficiency and microsecond lifetime. *Science* **363**, 601–606 (2019).

17. Soleilhavoup, M. & Bertrand, G. Cyclic (alkyl)(amino)carbenes (CAACs): Stable carbenes on the rise. *Acc. Chem. Res.* **48**, 256–266 (2015).

18. Melaimi, M., Jazzar, R., Soleilhavoup, M. & Bertrand, G. Cyclic (Alkyl)(amino)carbenes (CAACs): Recent Developments. *Angew. Chemie - Int. Ed.* **56**, 10046–10068 (2017).

19. Di, D. *et al.* High-performance light-emitting diodes based on carbene-metal-amides. *Science* **356**, 159–163 (2017).

20. Conaghan, P. J. *et al.* Efficient Vacuum-Processed Light-Emitting Diodes Based on Carbene–Metal–Amides. *Adv. Mater.* **30**, 1802285 (2018).

21. Romanov, A. S. *et al.* Dendritic Carbene Metal Carbazole Complexes as Photoemitters for Fully Solution-Processed OLEDs. *Chem. Mater.* acs.chemmater.8b05112 (2019). doi:10.1021/acs.chemmater.8b05112

22. Thompson, S., Eng, J. & Penfold, T. J. The intersystem crossing of a cyclic (alkyl)(amino) carbene gold (i) complex. *J. Chem. Phys.* (2018). doi:10.1063/1.5032185

23. Taffet, E. J., Olivier, Y., Lam, F., Beljonne, D. & Scholes, G. D. Carbene-Metal-Amide Bond Deformation, Rather Than Ligand Rotation, Drives Delayed Fluorescence. *J. Phys. Chem. Lett.* (2018). doi:10.1021/acs.jpclett.8b00503

24. Föller, J. & Marian, C. M. Rotationally Assisted Spin-State Inversion in Carbene-Metal-Amides Is an Artifact. *J. Phys. Chem. Lett.* **8**, 5643–5647 (2017).

25. Hamze, R. *et al.* ‘Quick-Silver’ from a Systematic Study of Highly Luminescent, Two-Coordinate, d10 Coinage Metal Complexes. *J. Am. Chem. Soc.* **141**, 8616–8626 (2019).

26. Cardona, C. M., Li, W., Kaifer, A. E., Stockdale, D. & Bazan, G. C. Electrochemical considerations for determining absolute frontier orbital energy levels of conjugated polymers for solar cell applications. *Adv. Mater.* **23**, 2367–2371 (2011).

## SUPPORTING INFORMATION

### Experimental section.

**General Considerations.** Unless stated otherwise, all reactions were carried out in air. Solvents were distilled and dried as required. Sodium *tert*-butoxide, 3-(*tert*-butyl)phenylboronic acid, were purchased from FluoroChem, SPhos Pd G2 was purchased from Sigma-Aldrich and used as received. The carbene ligand ($^{Ad}L$),[1,2,3] *N*-(2-chloro-4-(trifluoromethyl)phenyl)acetamide and 6-(*tert*-butyl)-3-(trifluoromethyl)-9H-carbazole carbazole,[4] and complexes ($^{Ad}L$)MCl (M = Cu and Au)[5] were obtained according to literature procedures. $^{1}H$ and $^{13}C\{^{1}H\}$ NMR spectra were recorded using a Bruker Avance DPX-300 MHz NMR spectrometer. $^{1}H$ NMR spectra (300.13 MHz) and $^{13}C\{^{1}H\}$ (75.47 MHz) were referenced to $CD_2Cl_2$ at δ 5.32 ($^{13}C$, δ 53.84), $C_6D_6$ at δ 7.16 ($^{13}C$, δ 128.4), $CDCl_3$ at δ 7.26 (δ $^{13}C$ 77.16) ppm. All electrochemical experiments were performed using an Autolab PGSTAT 302N computer-controlled potentiostat. Cyclic voltammetry (CV) was performed using a three-electrode configuration consisting of either a glassy carbon macrodisk working electrode (GCE) (diameter of 3 mm; BASi, Indiana, USA) combined with a Pt wire counter electrode (99.99 %; GoodFellow, Cambridge, UK) and an Ag wire pseudo-reference electrode (99.99 %; GoodFellow, Cambridge, UK). The GCE was polished between experiments using alumina slurry (0.3 μm), rinsed in distilled water and subjected to brief sonication to remove any adhering alumina microparticles. The metal electrodes were then dried in an oven at 100 °C to remove residual traces of water, the GCE was left to air dry and residual traces of water were removed under vacuum. The Ag wire pseudoreference electrodes were calibrated to the ferrocene/ferrocenium couple in MeCN at the end of each run to allow for any drift in potential, following IUPAC recommendations.[6] All electrochemical measurements were performed at ambient temperatures under an inert Ar atmosphere in MeCN containing complex under study (0.14 mM) and supporting electrolyte [n-$Bu_4N$][$PF_6$] (0.13 mM). Data were recorded with Autolab NOVA software (v. 1.11). Elemental analyses were performed by London Metropolitan University. UV-visible absorption spectra were recorded using a Perkin-Elmer Lambda 35 UV/vis spectrometer. Mass spectrometry data was obtained using APCI(ASAP) (Atmospheric Solids Analysis Probe) at the National Mass Spectrometry Facility at Swansea University.

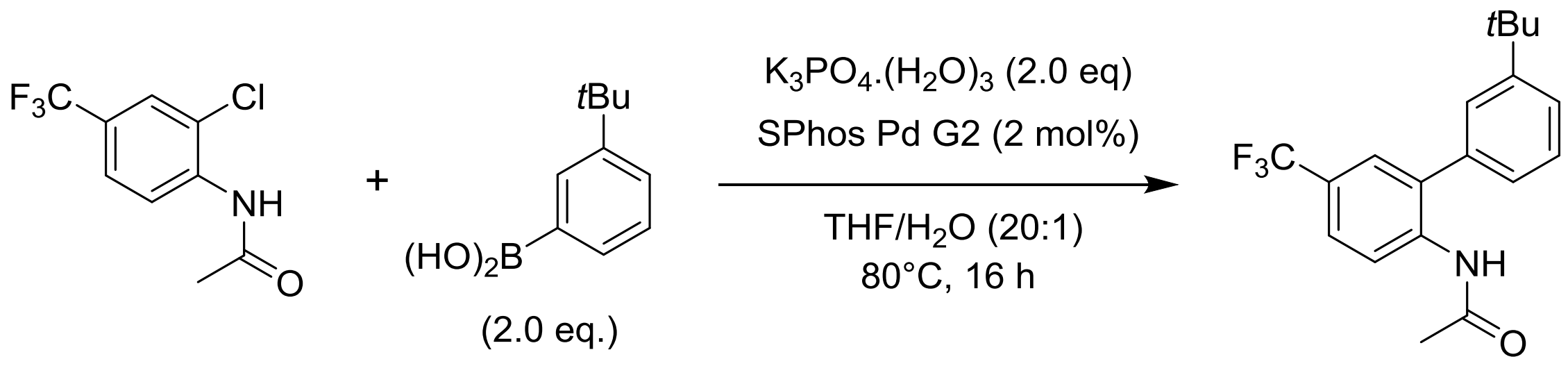

**92% (1.55 g)**

**Synthesis of *N*-(3'-(*tert*-butyl)-5-(trifluoromethyl)-[1,1'-biphenyl]-2-yl)acetamide:** *N*-(2-chloro-4-(trifluoromethyl)phenyl)acetamide (1 eq., 5.05 mmol, 1.20 g), 3-(*tert*-butyl)phenylboronic acid (2.0 eq., 10.1 mmol, 1.80 g,) and potassium phosphate trihydrate (2.0 eq., 10.1 mmol, 2.33 g) were mixed in THF/$H_2O$ 20:1 (10 mL) and purged with argon. SPhos Pd G2 (1 mol%, 0.051 mmol, 37 mg) was added and the mixture was heated at 80°C for 16 h. Reaction was cooled to r.t., $Et_2O$ (30 mL) was added and the mixture was filtered through Celite®. The filtrate was diluted with AcOEt (100 mL), washed with water and brine, and dried with $MgSO_4$. The solvent was evaporated and the residue was purified by silica column chromatography (PE/AcOEt) to afford the product as an off-white solid (92%, 1.55 g).

$^1$H NMR (300 MHz, $CDCl_3$): δ 8.53 (d, *J* = 8.7 Hz, 1H), 7.61 (pseudo dd, *J* = 8.7, 2.2 Hz, 1H), 7.52 – 7.42 (m, 3 x 1H overlapped), 7.40 – 7.37 (m, 1H), 7.37 – 7.32 (bs, NH), 7.19 (pseudo dt, *J* = 6.6, 1.9 Hz, 1H), 2.05 (s, 3H, Ac), 1.37 (s, 9H, *t*Bu). $^{13}$C NMR (75 MHz, $CDCl_3$) δ 168.4 (s, C=O), 152.7 (s, $\underline{C}$–*t*Bu), 138.0 (s, $C_q$), 136.4 (s, $C_q$), 132.2 (s, $C_q$)), 129.5 (s, CH), 127.1 (q, *J* = 3.6 Hz, $\underline{C}$H–C–$CF_3$), 126.4 (s, CH), 126.3 (s, CH), 125.8 (s, CH), 125.5 (q, *J* = 3.7 Hz, $\underline{C}$H–C–$CF_3$), 124.2 (q, *J* = 271.8 Hz, $CF_3$), 120.8 (s, CH), 35.1 (s, $\underline{C}(CH_3)_3$), 31.5 (s, $C(\underline{C}H_3)_3$), 25.0 (s, $CH_3$ Ac), ($\underline{C_{ipso}}$–$CF_3$ was not observed due to overlap with aromatic signals). $^{19}$F NMR (282 MHz, $CDCl_3$) δ -62.1 ppm. Anal. Calcd. for $C_{19}H_{20}F_3NO$ 335.37): C, 68.05; H, 6.01; N, 4.18. Found: C, 68.17; H, 6.18; N, 4.31.

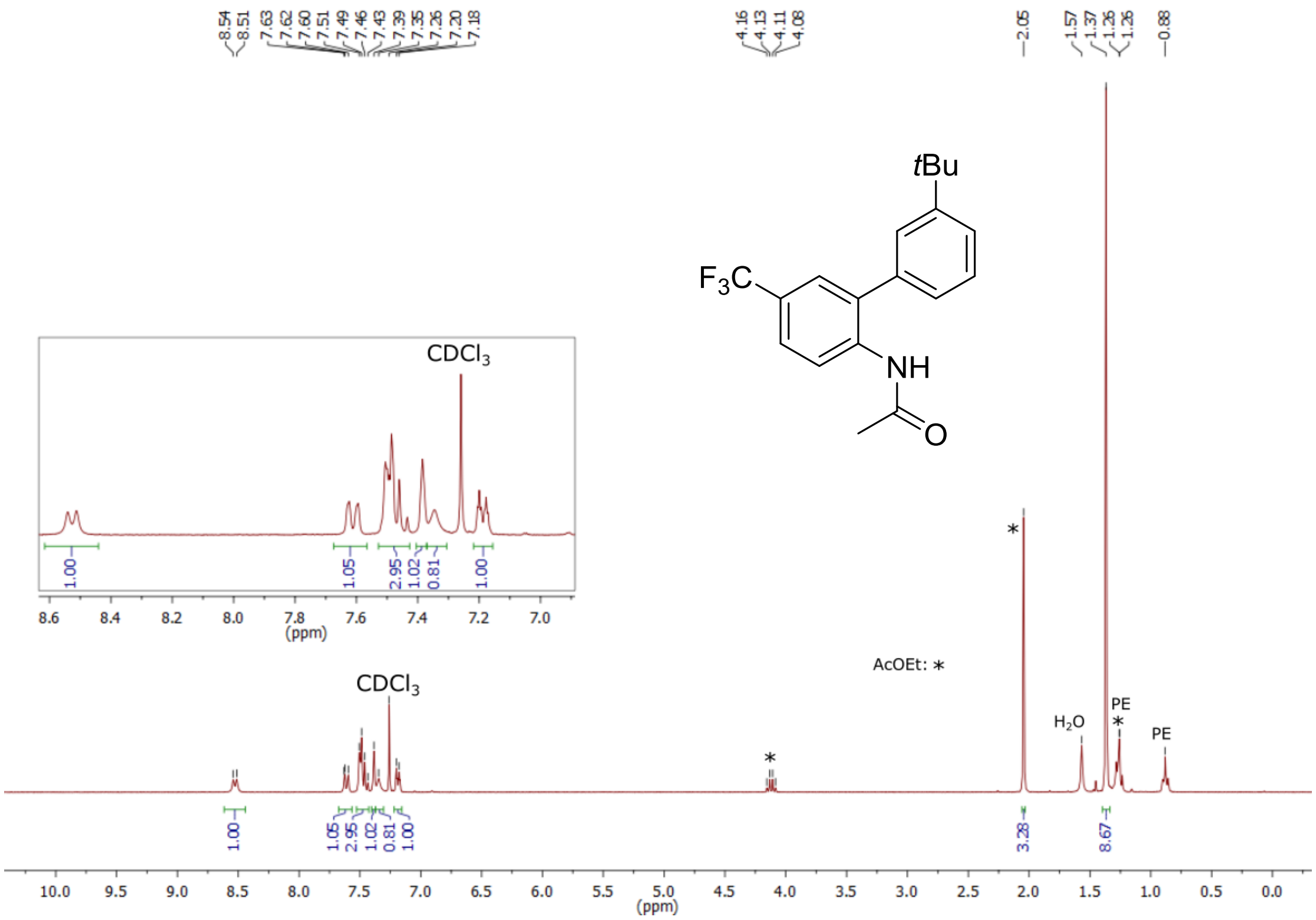

$^{1}$H NMR (300 MHz, $CDCl_3$)

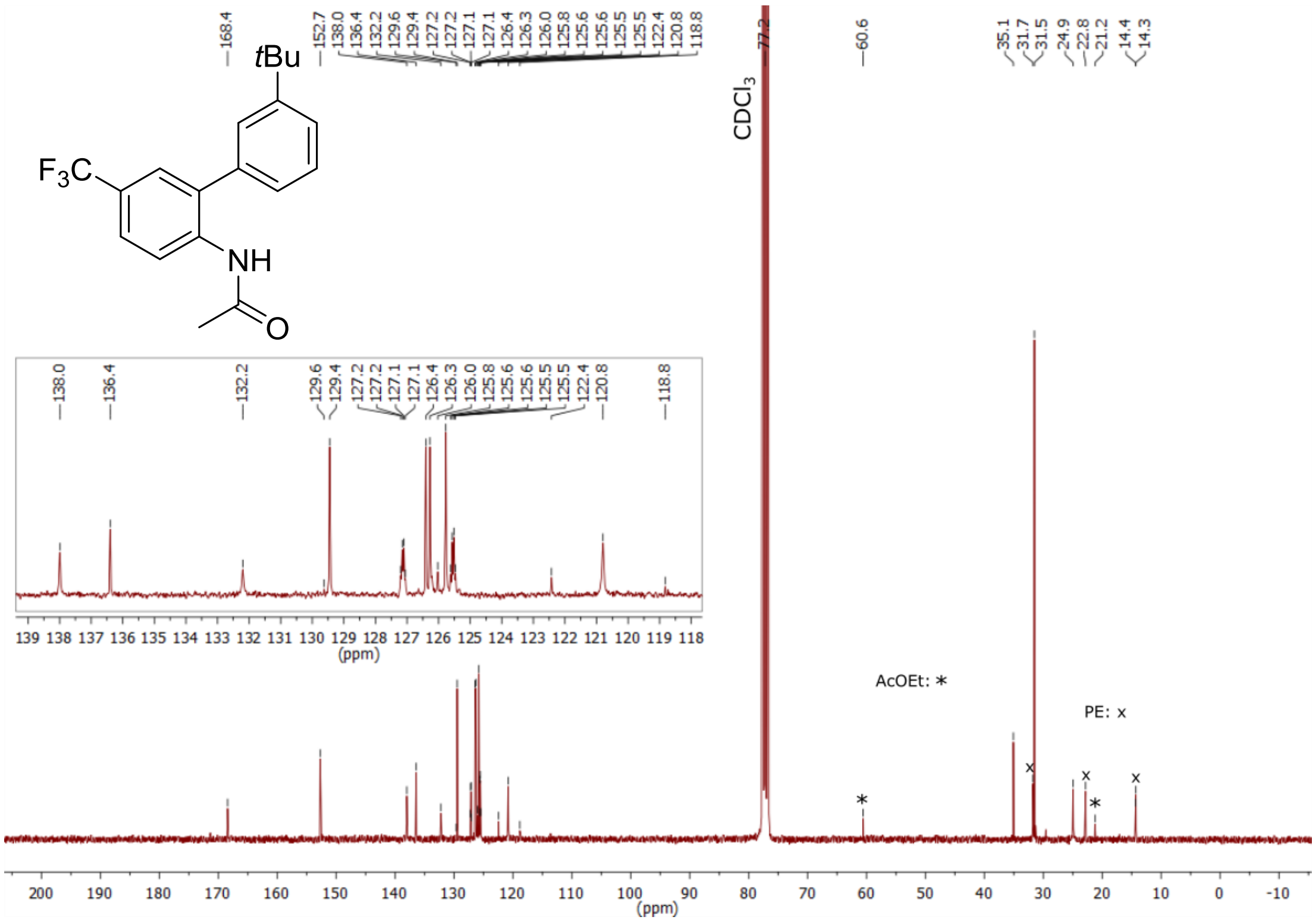

$^{13}$C NMR (75 MHz, $CDCl_3$)

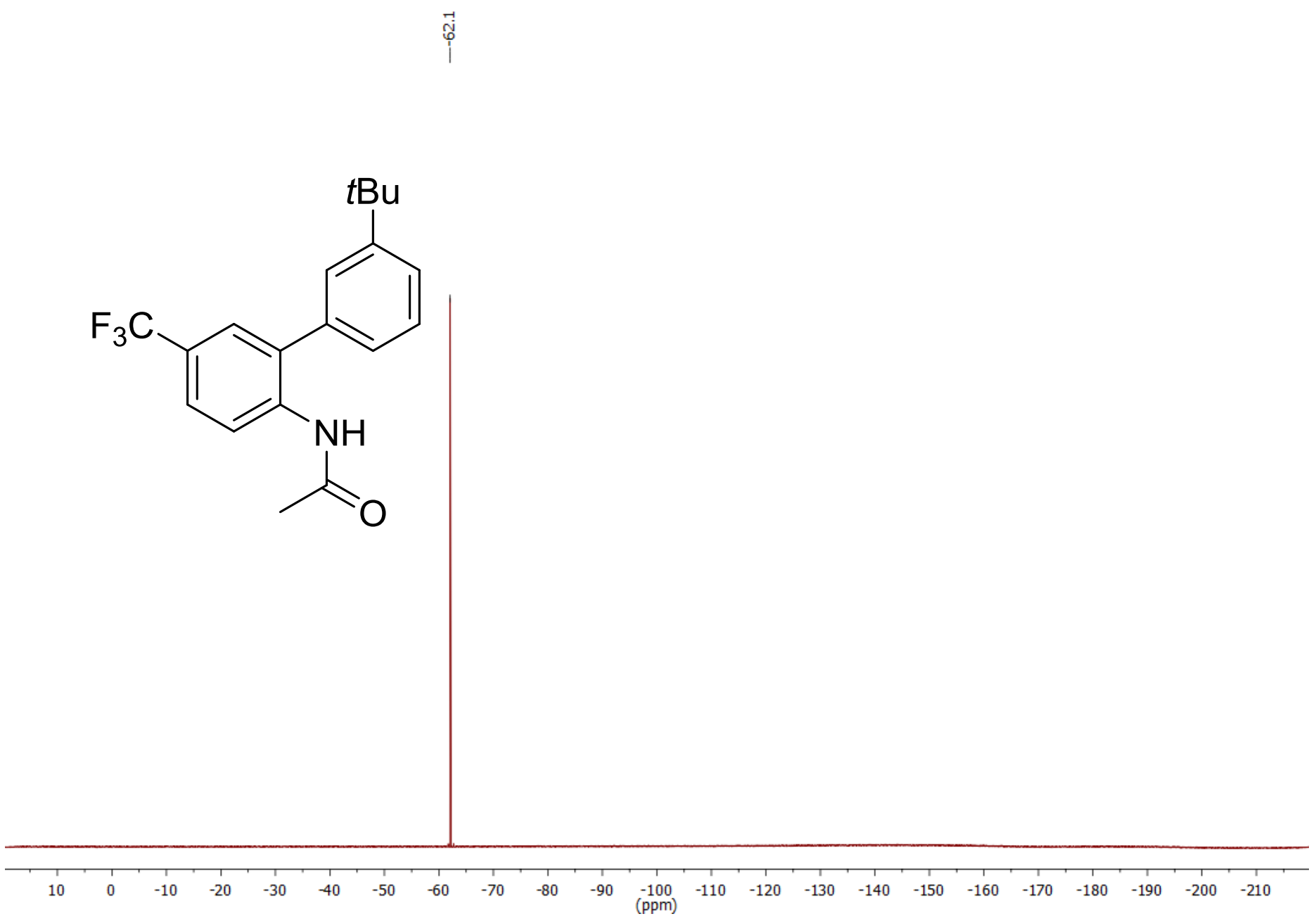

$^{19}$F NMR (282 MHz, $CDCl_3$)

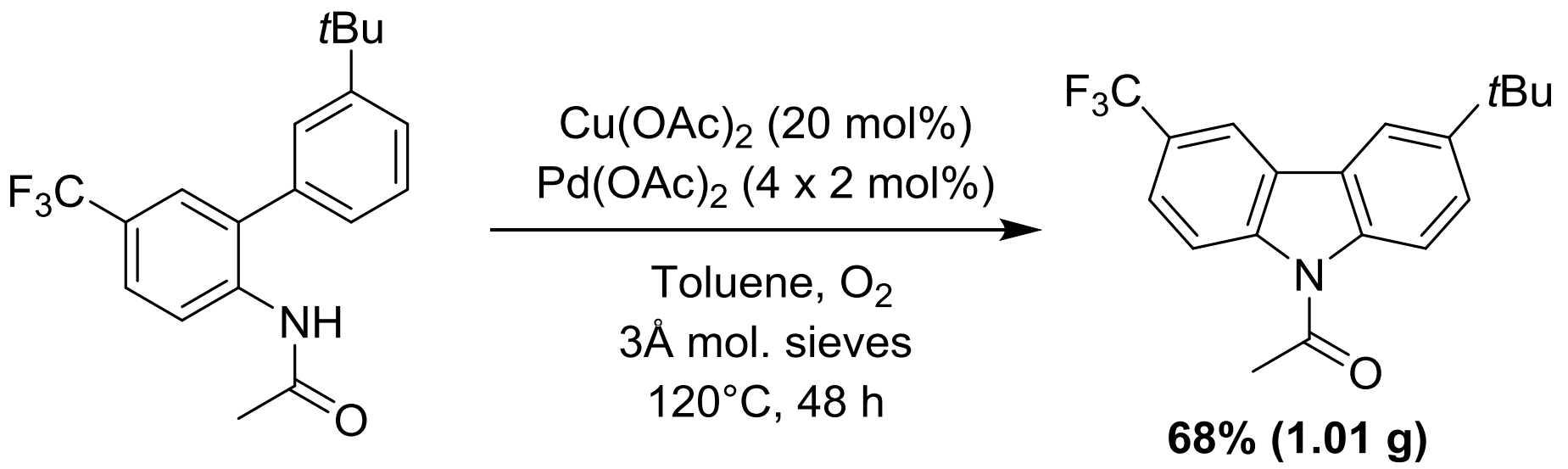

**Synthesis of 6-(*tert*-butyl)-3-(trifluoromethyl)-9-acetylcarbazole:** In an oven-dried Schlenk tube, under argon, *N*-(3'-(*tert*-butyl)-5-(trifluoromethyl)-[1,1'-biphenyl]-2-yl)acetamide (1 eq., 4.47 mmol, 1.50 g), $Cu(OAc)_2$ (20 mol%, 0.89 mmol, 162 mg), $Pd(OAc)_2$ (2 mol%, 0.089 mmol, 20 mg) and 3 Å molecular sieves were mixed in toluene (20 mL). The flask was purged by oxygen and heated at 120°C. Reaction completion was followed and a portion of $Pd(OAc)_2$ was added each day (3 x 2 mol%, 3 x 20 mg). After 2 days, reaction was completed. The mixture was cooled to r.t., diluted with AcOEt (60 mL) and filtered through Celite®. The filtrate was diluted with AcOEt (100 mL) washed with water and brine and dried with $MgSO_4$. The solvent was evaporated and the residue was purified by silica column chromatography (PE/AcOEt) to afford the product as an off-white solid (68%, 1.01 g).

$^1$H NMR (300 MHz, $CDCl_3$): δ 8.46 (d, $J$ = 8.8 Hz, 1H, $CH^1$), 8.27 (pseudo s, 1H, $CH^4$), 8.04 (d, $J$ = 2.1 Hz, 1H, $CH^5$), 8.01 (d, $J$ = 8.9 Hz, 1H, $CH^8$), 7.71 (pseudo dd, $J$ = 8.8, 1.9 Hz, 1H, $CH^2$),

7.60 (dd, $J$ = 8.9, 2.1 Hz, 1H, $CH^{7}$), 2.90 (s, 3H, Ac), 1.45 (s, 9H, *t*Bu). $^{13}$C NMR (75 MHz, $CDCl_3$) δ 170.1 (s, C=O), 147.5 (C–*t*Bu), 140.9 (s, $C_q$), 137.1 (s, $C_q$), 126.8 (s, $C_q$), 126.1 (s, $CH^{7}$), 126.0 (q, $J$ = 32.6 Hz, C–$CF_3$), 125.6 (s, $C_q$), 124.7 (q, $J$ = 271.7 Hz, $CF_3$), 124.2 (q, $J$ = 3.6 Hz, $CH^{2}$), 117.0 (s, $CH^{1}$ overlapped with $CH^{4}$), 116.9 (q, $J$ = 3.9 Hz, $CH^{4}$ overlapped with $CH^{1}$), 116.8 (s, $CH^{5}$), 115.6 (s, $CH^{8}$), 34.9 (s, $C(CH_3)_3$, 31.8 ($C(CH_3)_3$, 27.8 (s, $CH_3$ Ac). $^{19}$F NMR (282 MHz, $CDCl_3$) δ -61.2. Anal. Calcd. for $C_{19}H_{18}F_3NO$ (333.35): C, 68.46; H, 5.44; N, 4.20. Found: C, 68.13; H, 5.68; N, 3.97.

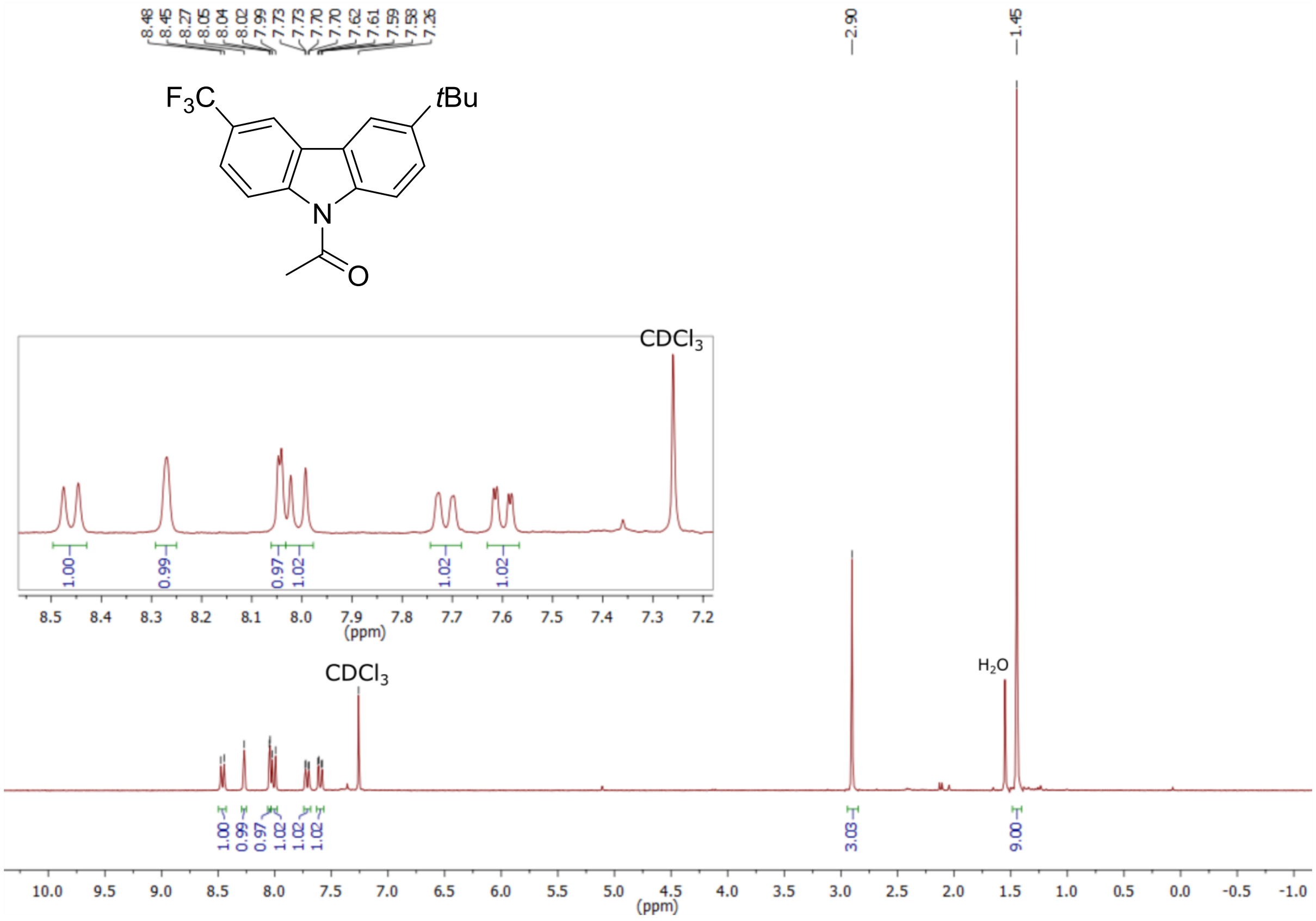

$^{1}$H NMR (300 MHz, $CDCl_3$)

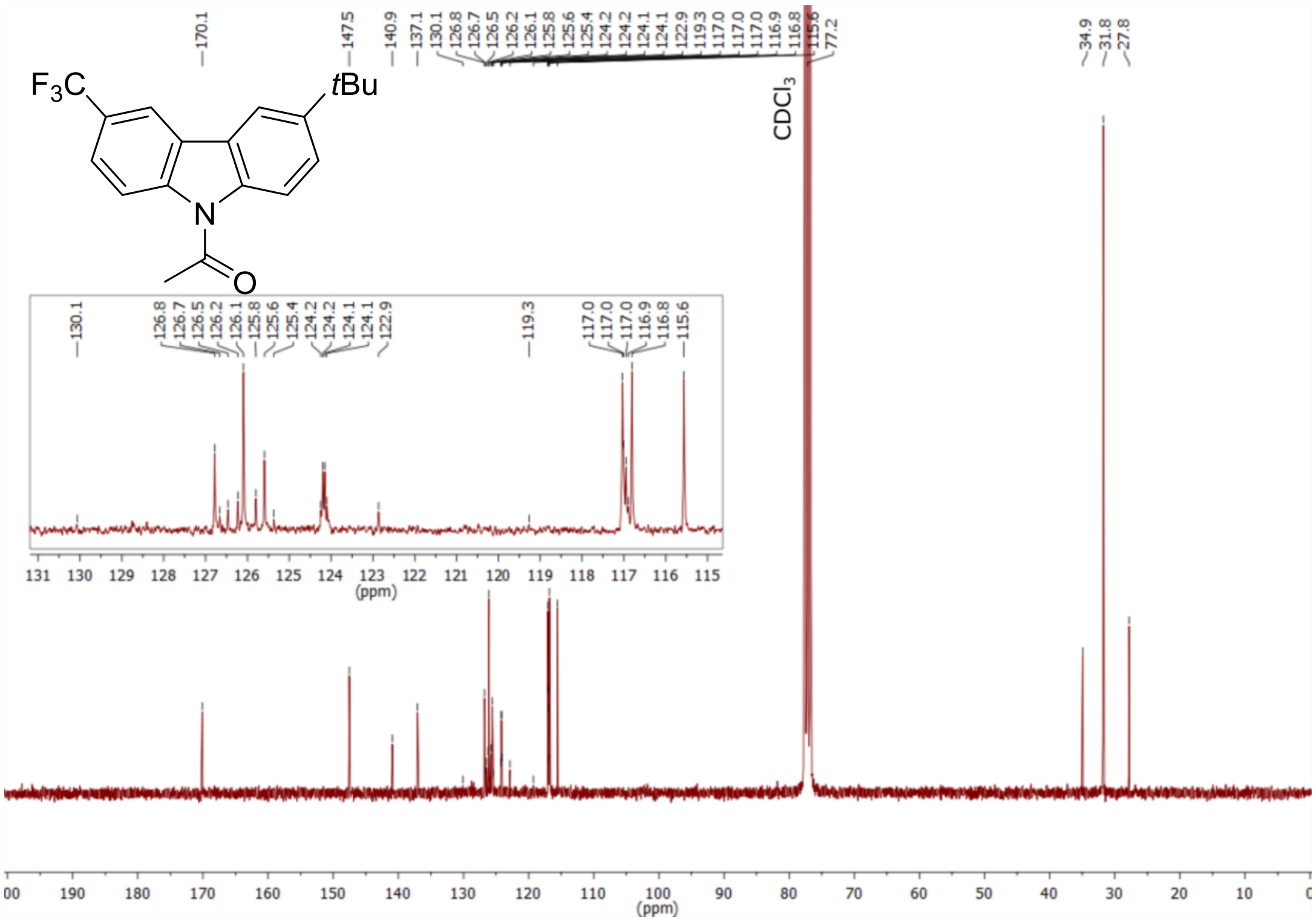

$^{13}C$ NMR (75 MHz, $CDCl_3$)

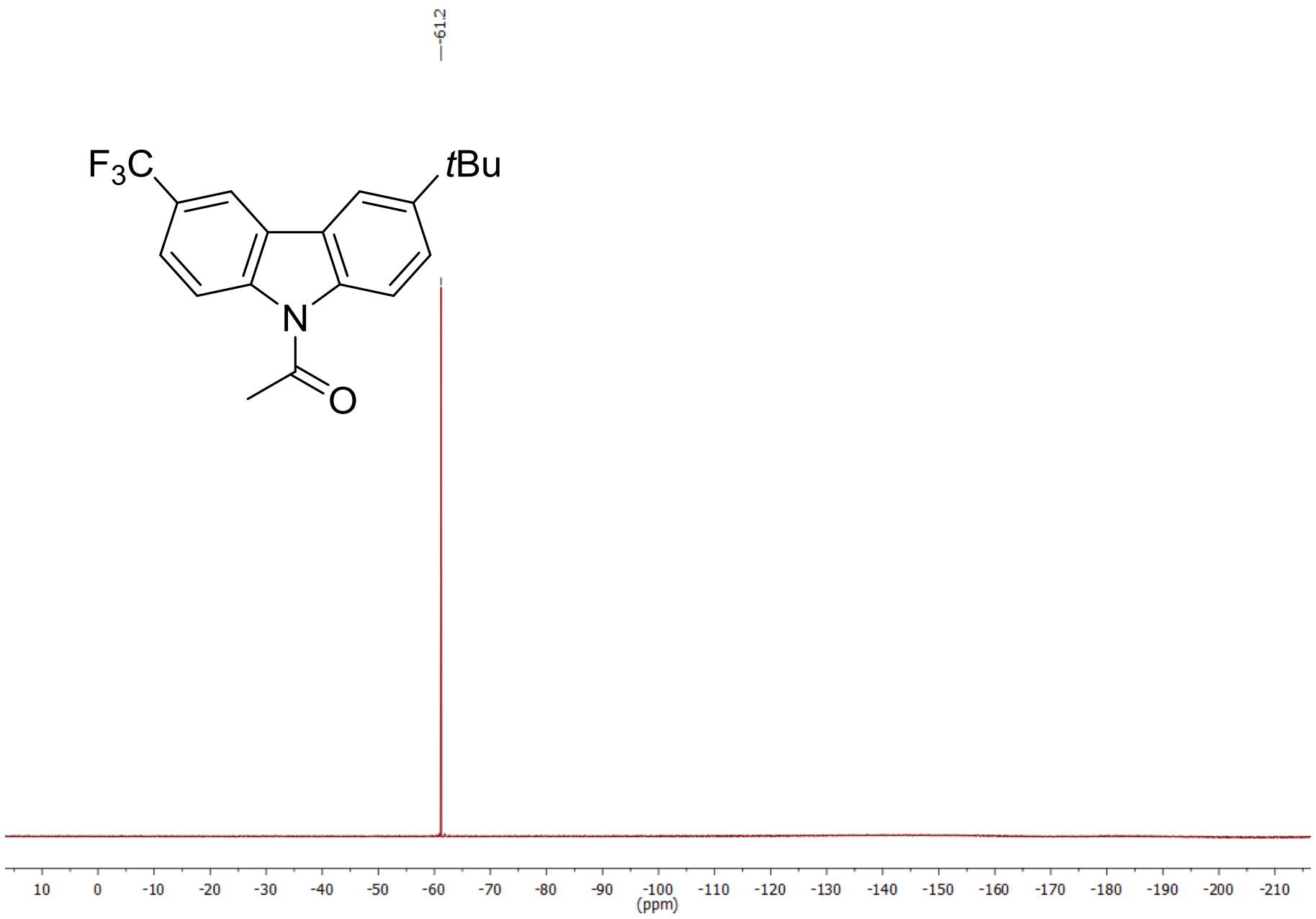

$^{19}F$ NMR (282 MHz, $CDCl_3$)

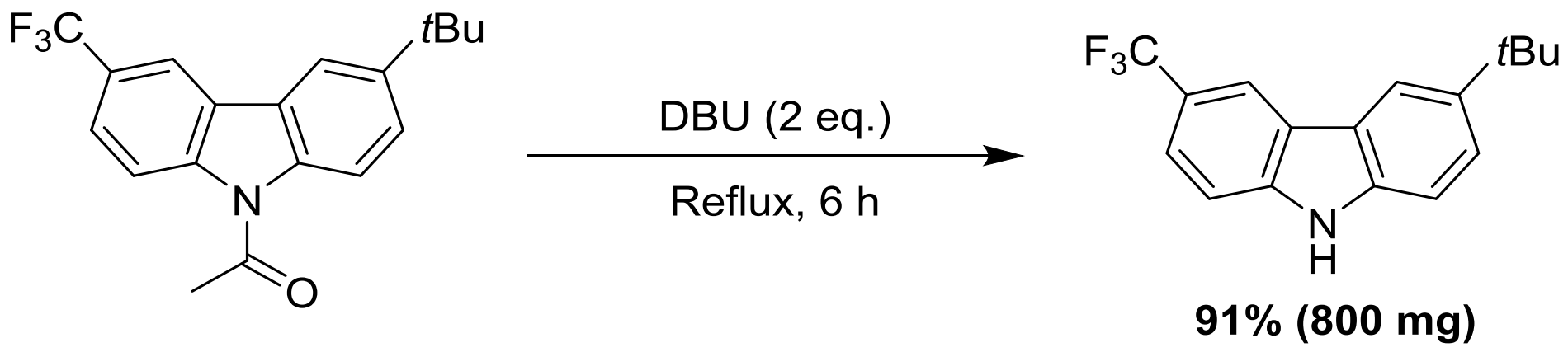

**Synthesis of 6-(*tert*-butyl)-3-(trifluoromethyl)-9H-carbazole.** DBU (2 eq., 6.06 mmol, 904 µL) was added to 6-(*tert*-butyl)-3-(trifluoromethyl)-9-acetylcarbazole (1 eq., 3.03 mmol, 1.01 g) in MeOH (30 mL), and the mixture was refluxed for 6 h. The reaction was cooled to r.t. and volatiles were evaporated. AcOEt (120 mL) was added, washed with water and brine, and dried over $MgSO_4$. The residue was purified by silica column chromatography (PE/AcOEt) to afford the product as a white solid (91%, 800 mg).

$^{1}H$ NMR (300 MHz, $CDCl_3$): δ 8.37 (s, 1H, $CH^4$), 8.16 (bs, 1H, NH), 8.12 (pseudo s, 1H, $CH^5$), 7.64 (pseudo dd, $J$ = 8.5, 1.7 Hz, 1H, $CH^2$), 7.56 (dd, $J$ = 8.6, 1.9 Hz, 1H, $CH^8$), 7.47 (d, $J$ = 8.5 Hz, 1H, $CH^1$), 7.41 (d, $J$ = 8.6 Hz, 1H, $CH^7$), 1.45 (s, 9H, *t*Bu). $^{13}C$ NMR (75 MHz, $CDCl_3$) δ 143.6 (s, $\underline{C}$–*t*Bu), 141.5 (s, $C_q$), 138.2 (s, $C_q$), 125.5 (q, $J$ = 271.2 Hz, $CF_3$), 125.0 (s, $CH^7$), 123.4 (s, $C_q$), 122.8 (s, $C_q$), 122.5 (q, $J$ = 3.7 Hz, $CH^2$), 121.7 (q, $J$ = 32.1 Hz, $\underline{C}$–$CF_3$), 117.9 (q, $J$ = 4.1 Hz, $CH^4$), 116.8 (s, $CH^5$), 110.7 (s, $CH^1$), 110.6 (s, $CH^8$), 34.9 (s, $\underline{C}(CH_3)_3$), 32.1 (s, $C(\underline{C}H_3)_3$). $^{19}F$ NMR (282 MHz, $CDCl_3$) δ -60.1. Anal. Calcd. for $C_{17}H_{16}F_3N$ (291.32): C, 70.09; H, 5.54; N, 4.81. Found: C, 69.82; H, 5.72; N, 4.63.

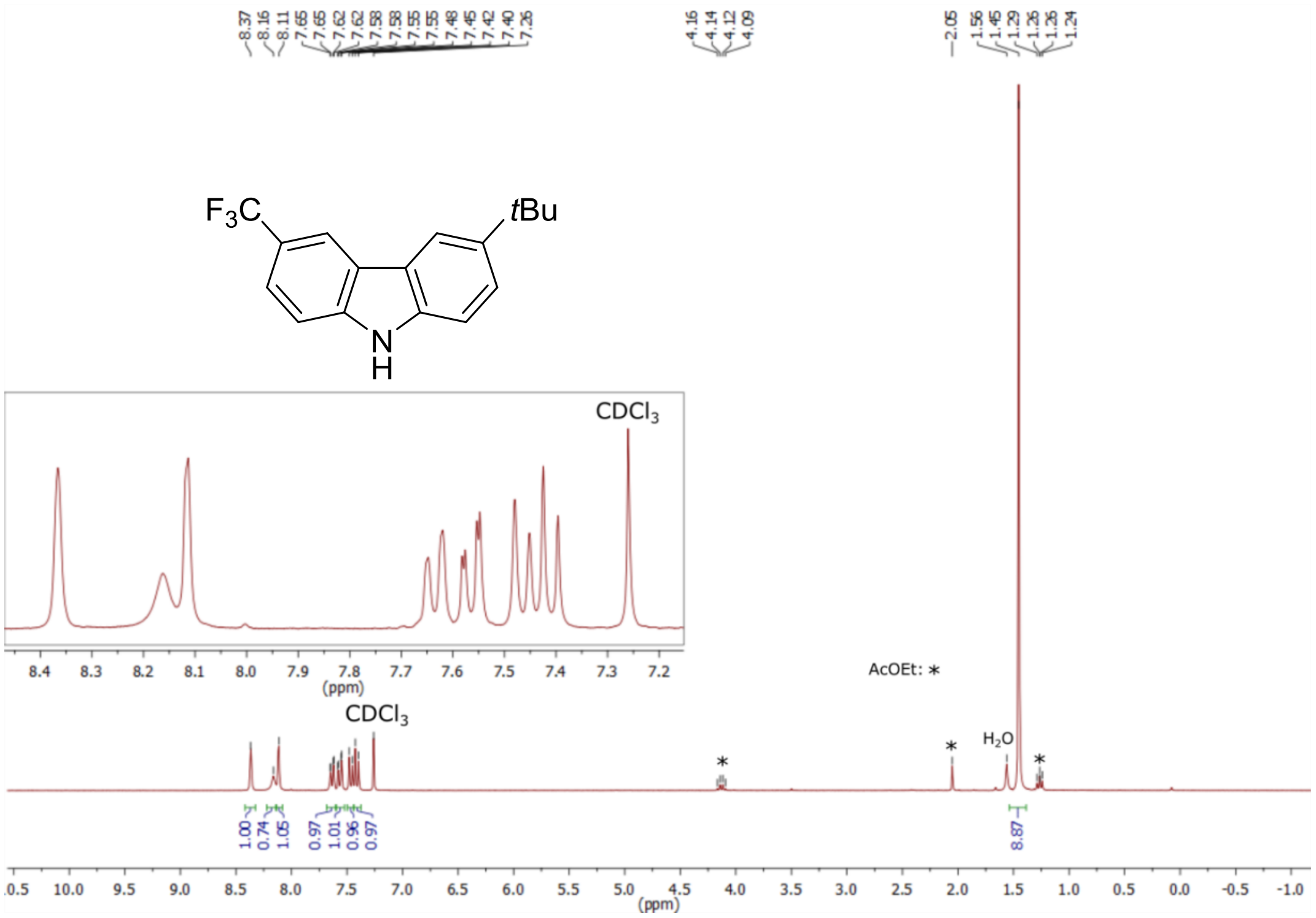

$^{1}H$ NMR (300 MHz, $CDCl_3$)

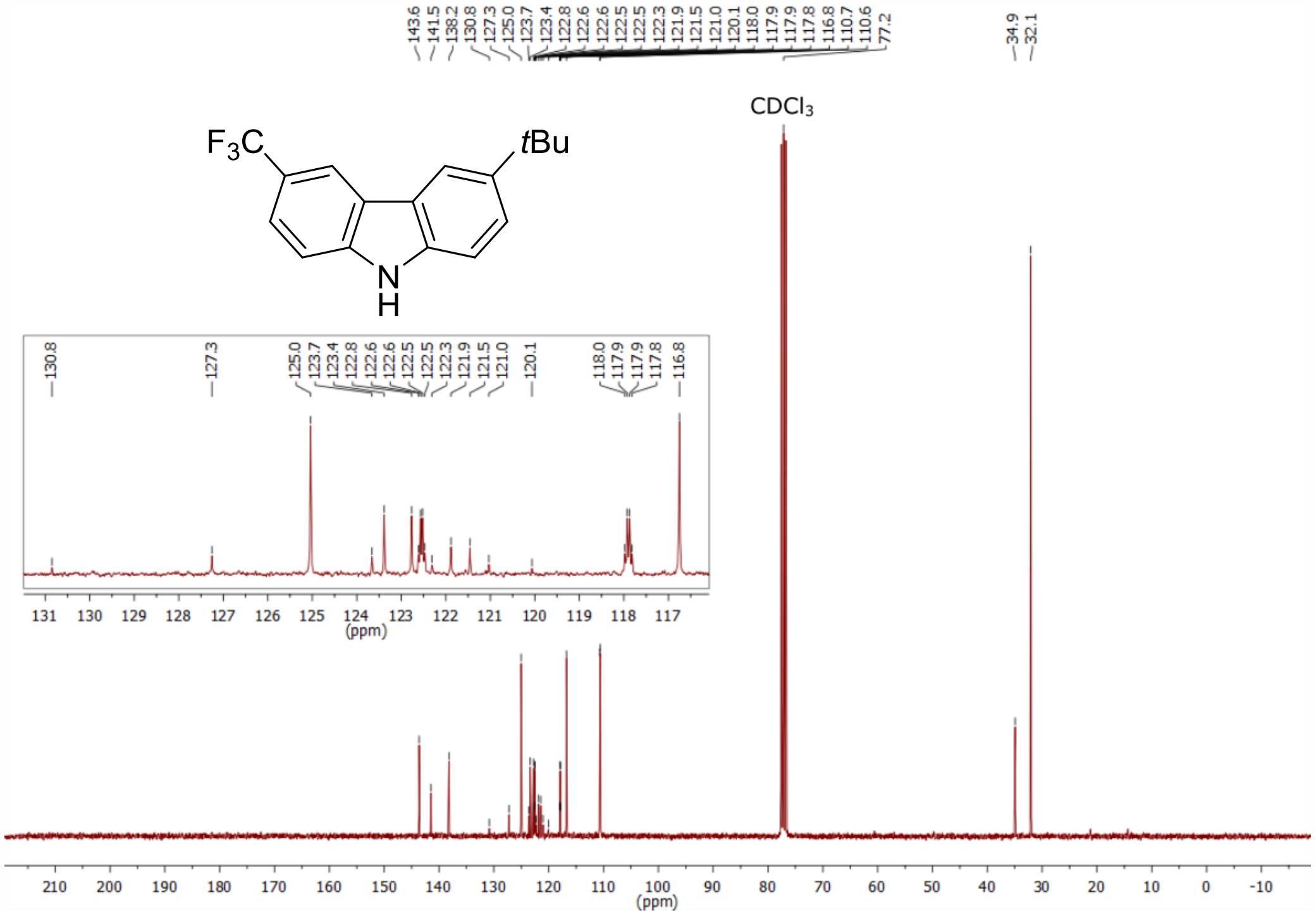

$^{13}C$ NMR (75 MHz, $CDCl_3$)

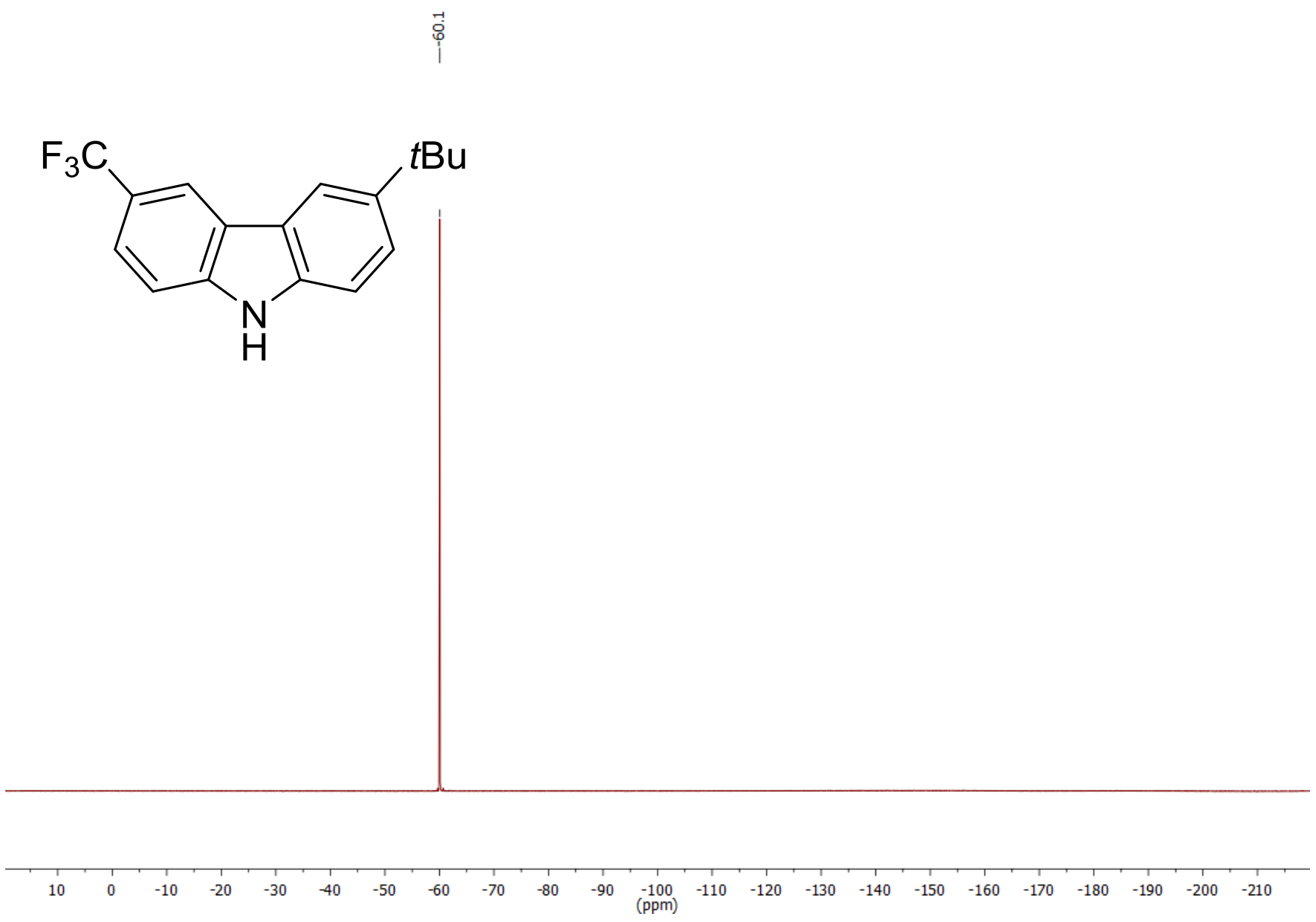

$^{19}$F NMR (282 MHz, $CDCl_3$)

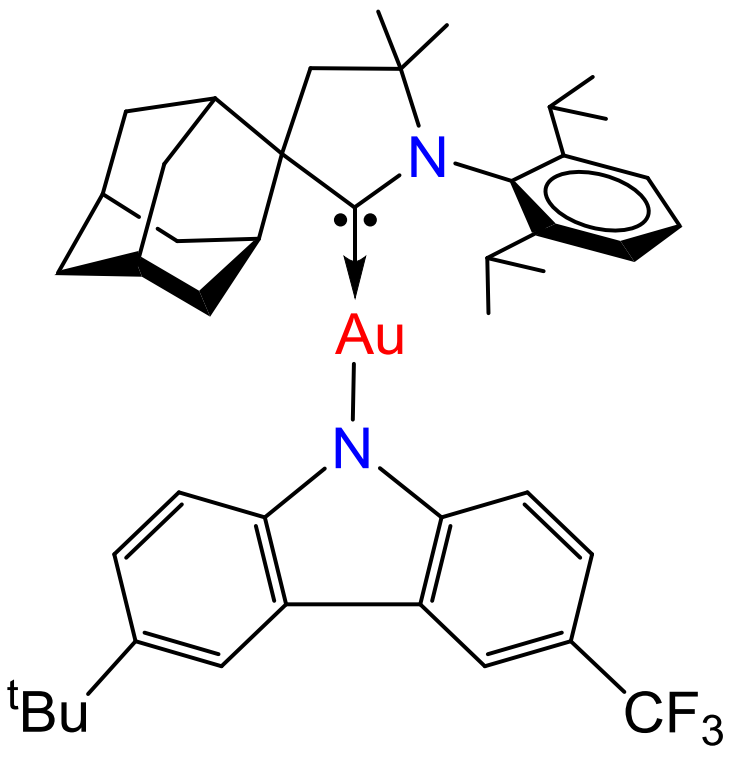

**Synthesis of ($^{Ad}$CAAC)Au(6-(*tert*-butyl)-3-(trifluoromethyl)-9H-carbazole) (1):** In a Schlenk tube, ($^{Ad}$L)AuCl (3.52 g, 5.77 mmol,), 6-(*tert*-butyl)-3-(trifluoromethyl)-9H-carbazole (1.68 g, 5.77 mmol), and $^t$BuONa (0.56 g, 5.83 mmol,) were stirred in THF (75 mL) for 6 h. The mixture was filtered through Celite. The filtrate was concentrated and washed with hexane to afford the product as a white solid. Yield: 93% (4.65 g, 5.37 mmol).

$^1$H NMR (300 MHz, $CD_2Cl_2$): δ 8.20 (s, 1H, CH$^4$ Cz), 7.99 (d, *J* = 2.0 Hz, 1H CH$^5$ Cz), 7.71 (t, *J* = 7.8 Hz, 1H, *p*-CH Dipp), 7.47 (d, *J* = 7.8 Hz, 2H, *m*-CH Dipp), 7.30 (dd, *J* = 8.6, 2.0 Hz, 1H, CH$^7$ Cz), 7.25 (dd, *J* = 8.6, 1.5 Hz 1H, CH$^2$ Cz), 6.87 (d, *J* = 8.6 Hz, 1H, CH$^8$ Cz), 6.42 (d, *J* = 8.6 Hz, 1H, CH$^1$ Cz), 4.31 (d, *J* = 12.9 Hz, 2H, $CH_2$ Adamantyl), 2.90 (Sept, *J* = 6.7 Hz, 2H, CH *i*Pr Dipp), 2.45 (pseudo s, 2H + 1H, $CH_2$ CAAC overlapping with CH Adamantyl), 2.19 – 1.86 (m, 11H, Adamantyl), 1.44 (s, 6H, $C(CH_3)_2$ CAAC), 1.39 (s, 9H, *t*Bu), 1.36 – 1.30 (m, 12H, $CH_3$ *i*Pr Dipp). $^{13}$C NMR (75 MHz, $CD_2Cl_2$) δ 244.1 (s, C: CAAC), 151.9 (s, $C_q$ Cz), 148.8 (s, $C_q$ Cz), 146.2 (s, *o*-C Dipp), 140.4 (s, C–*t*Bu), 136.7 (s, *i*-C Dipp), 130.0 (s, *p*-CH Dipp), 26.8 (q, *J* = 270.3

Hz, $CF_3$), 123.9 (s, $C_q$ Cz), 123.8 (s, $C_q$ Cz), 123.0 (s, $CH^7$ Cz), 119.8 (q, $J$ = 3.2 Hz, $CH^2$ Cz), 117.1 (q, $J$ = 31.3 Hz, $\underline{C}$–$CF_3$ overlapping with $CH^4$ Cz), 116.9 (q, $J$ = 4.2 Hz, $CH^4$ Cz overlapping with C–$CF_3$), 115.9 (s, $CH^5$ Cz), 114.0 (s, $CH^1$ Cz), 113.9 (s, $CH^8$ Cz), 77.6 (s, s, $\underline{C}(CH_3)_2$ CAAC), 64.5 (s, $\underline{C}$–C: CAAC), 49.1 (s, $CH_2$ CAAC), 39.4 (s, Adamantyl), 37.6 (s, Adamantyl), 35.8 (s, Adamantyl), 34.8 (s, 2 C overlapped, $\underline{C}(CH_3)_3$ and CH Adamantyl) 32.2 (s, $C(\underline{C}H_3)_3$) 29.6 m, 2 C overlapped, $C(\underline{C}H_3)_2$ and CH *i*Pr Dipp), 28.6 (s, Adamantyl), 27.7 (s, Adamantyl), 26.5 (s, $CH_3$ *i*Pr Dipp), 23.4 (s, $CH_3$ *i*Pr Dipp). $^{19}F$ NMR (282 MHz, $CD_2Cl_2$) δ -59.1. Anal. Calcd. for $C_{44}H_{54}AuF_3N_2$ (864.89): C, 61.10; H, 6.29; N, 3.24. Found: C, 61.35; H, 6.07; N, 3.43. $C_{44}H_{54}AuF_3N_2H$ theoretical $[M+H^+]$ = 865.3983, HRMS (APCI(ASAP)) = 865.3997

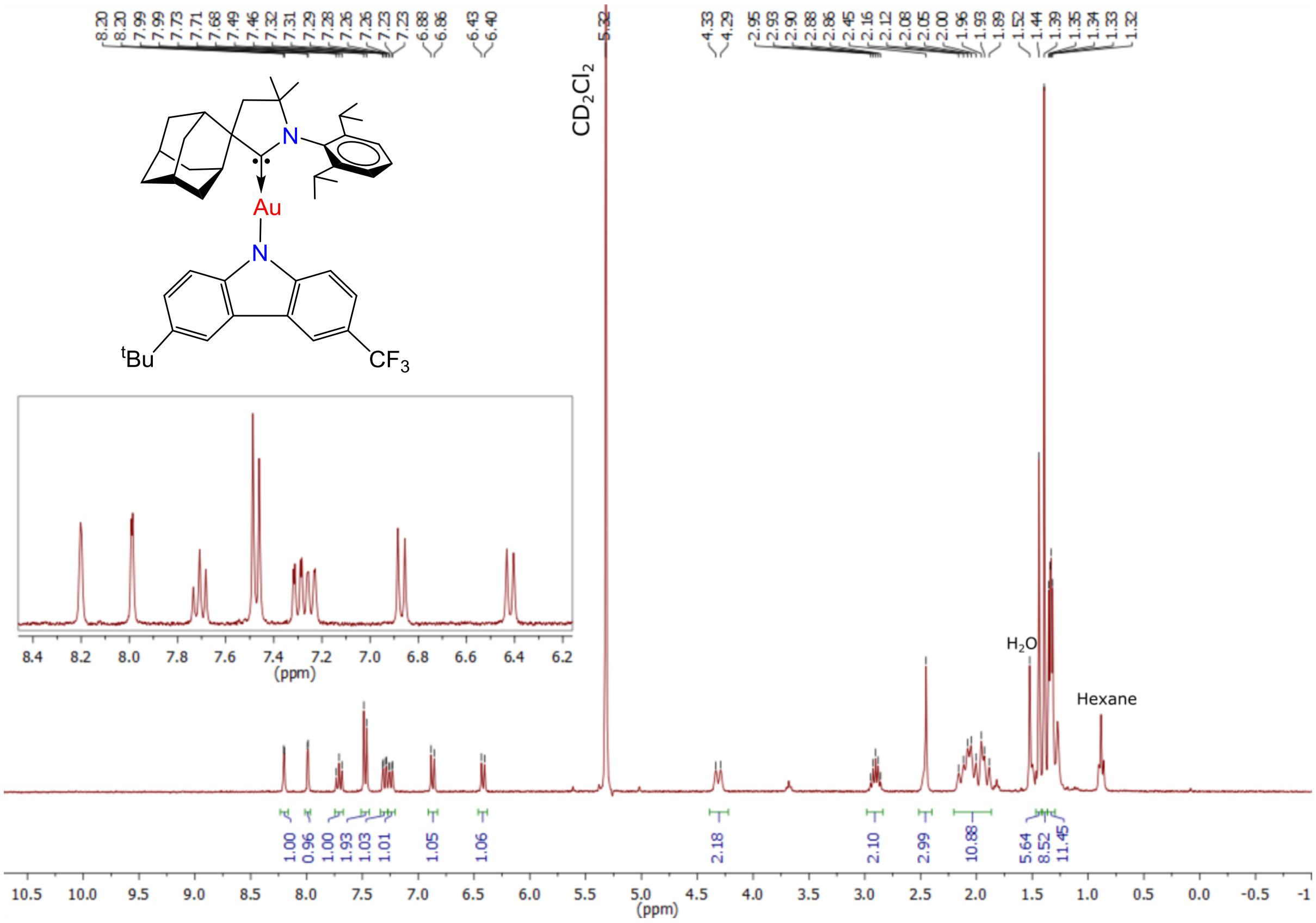

$^1H$ NMR (300 MHz, $CD_2Cl_2$)

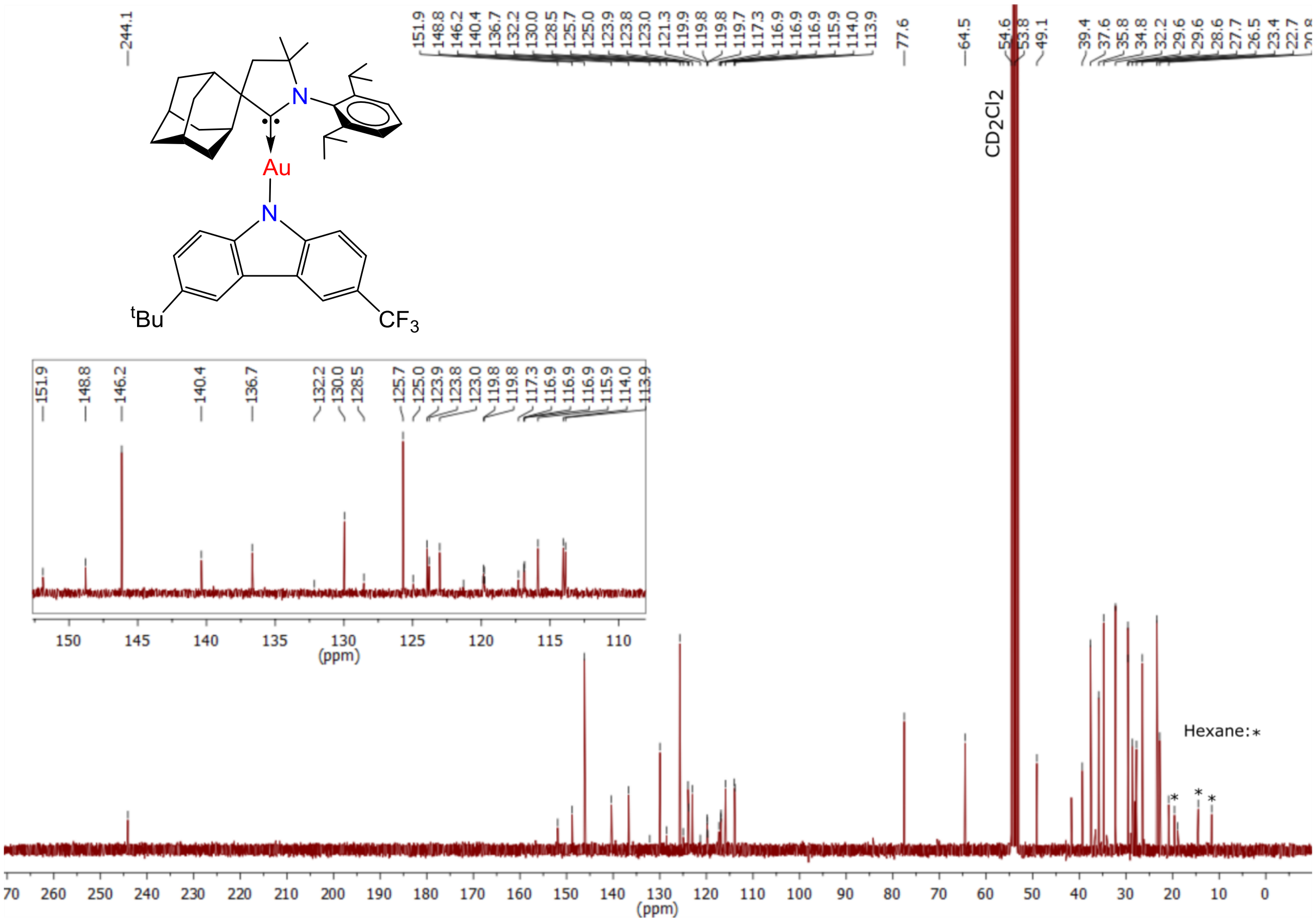

$^{13}C$ NMR (75 MHz, $CD_2Cl_2$)

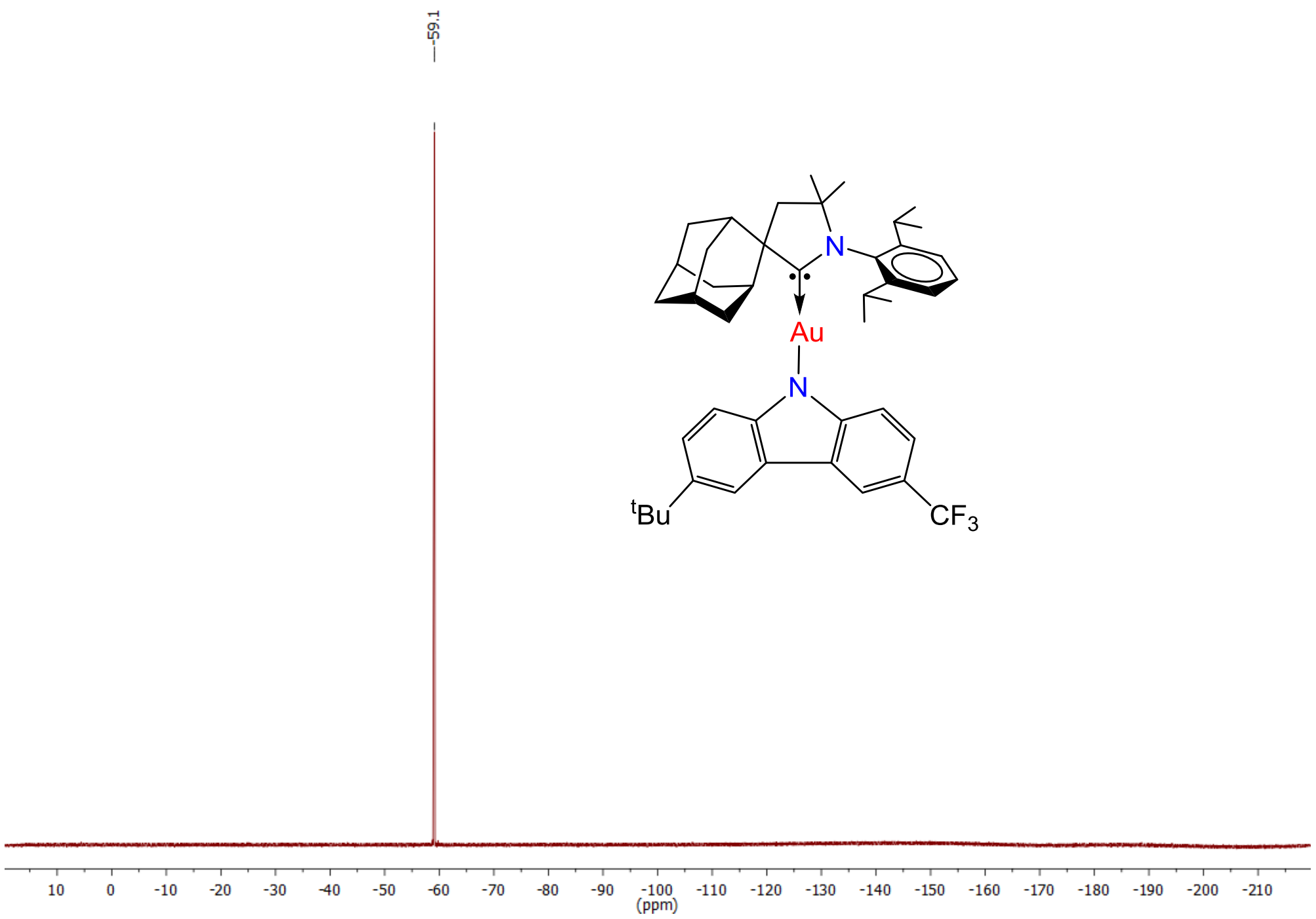

$^{19}F$ NMR (282 MHz, $CDCl_3$)

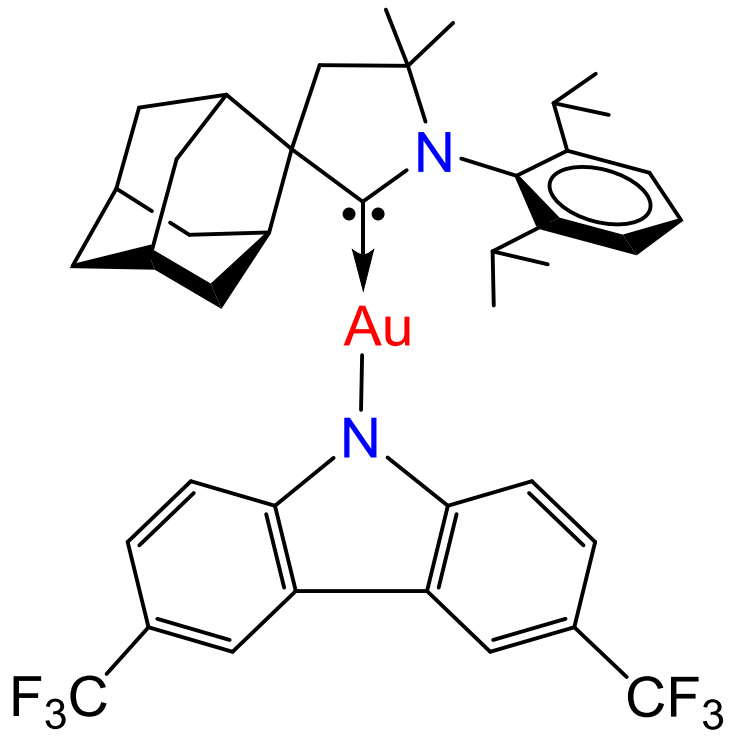

**Synthesis of ($^{Ad}$CAAC)Au(3,6-bis(trifluoromethyl)-9H-carbazole) (2):**

In a Schlenk tube, ($^{Ad}$L)AuCl (0.78 g, 1.28 mmol), 3,6-bis(trifluoromethyl)-9H-carbazole (0.373 g 1.28 mmol), and $^{t}$BuONa (0.140 mg 1.45 mmol) were stirred in THF (40 mL) for 6 h. The mixture was filtered through Celite. The filtrate was concentrated and washed with hexane to afford the product as a white solid. Yield: 85% (0.95 g, 1.08 mmol).

$^{1}$H NMR (300 MHz, $CD_2Cl_2$): δ 8.27 (s, 2H, $CH^4$ Cz), 7.72 (t, *J* = 7.8 Hz, 1H, *p*-CH Dipp), 7.48 (d, *J* = 7.8 Hz, 2H, *m*-CH Dipp), 7.40 (d, *J* = 8.6 Hz, 2H, $CH^2$ Cz), 6.73 (d, *J* = 8.6 Hz, 2H, $CH^1$ Cz), 4.27 (d, *J* = 12.7 Hz, 2H, $CH_2$ Adamantyl), 2.89 (sept, *J* = 6.6 Hz, 2H, CH *i*Pr Dipp), 2.44 (pseudo s, 2H + 1H, $CH_2$ CAAC overlapping with CH Adamantyl), 2.20–1.99 (m, 7H, Adamantyl), 1.97 – 1.85 (m, 4H, Adamantyl), 1.44 (s, 6H, $C(CH_3)_2$ CAAC), 1.40–1.26 (m, 12H, $CH_3$ *i*Pr Dipp). $^{13}$C NMR (75 MHz, $CD_2Cl_2$) δ 243.5 (s, C: CAAC), 152.3 (s, $C_q$ Cz), 146.2 (s, *o*-C Dipp), 136.7 (s, *i*-C Dipp), 130.1 (s, *p*-CH Dipp), 126.4 (q, *J* = 270.7 Hz, $CF_3$), 125.8 (s, *m*-CH Dipp), 123.6 (s, $C_q$ Cz), 121.3 (q, *J* = 3.3 Hz, $CH^2$ Cz), 118.8 (q, *J* = 31.5 Hz, $\underline{C}$–$CF_3$), 117.5 (q, *J* = 4.1 Hz, $CH^4$ Cz), 114.7 (s, $CH^1$ Cz), 77.8 (s, $\underline{C}(CH_3)_2$ CAAC), 64.5 (s, $\underline{C}$–C: CAAC), 49.0 (s, $CH_2$ CAAC), 39.3 (s, Adamantyl), 37.6 (s, Adamantyl), 35.9 (s, Adamantyl), 34.7 (s, Adamantyl), 29.6 (m, 2 C overlapped, $C(\underline{C}H_3)_2$ and CH *i*Pr Dipp), 28.6 (s, Adamantyl), 27.7 (s, Adamantyl), 26.5 (s, $CH_3$ *i*Pr Dipp), 23.3 (s, $CH_3$ *i*Pr Dipp). $^{19}$F NMR (282 MHz, $CD_2Cl_2$) δ -59.6. Anal. Calcd. for $C_{41}H_{45}AuF_6N_2$ (876.78): C, 56.17; H, 5.17; N, 3.20. Found: C, 55.83; H, 5.38; N, 3.02. $C_{41}H_{45}AuF_6N_2H$ theoretical $[M+H^+]$ = 877.3231, HRMS (APCI(ASAP)) = 877.3241

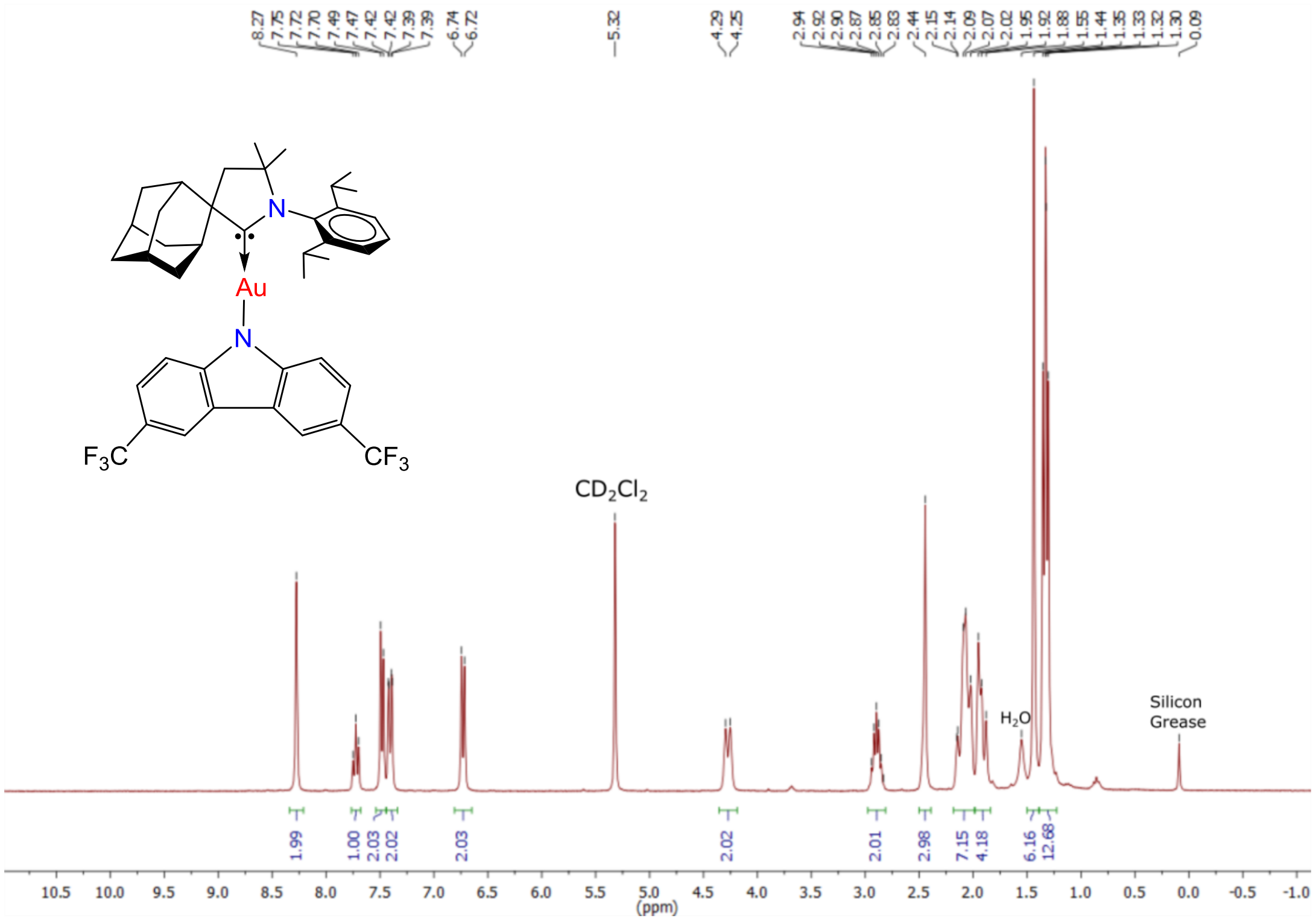

$^{1}H$ NMR (300 MHz, $CD_2Cl_2$)

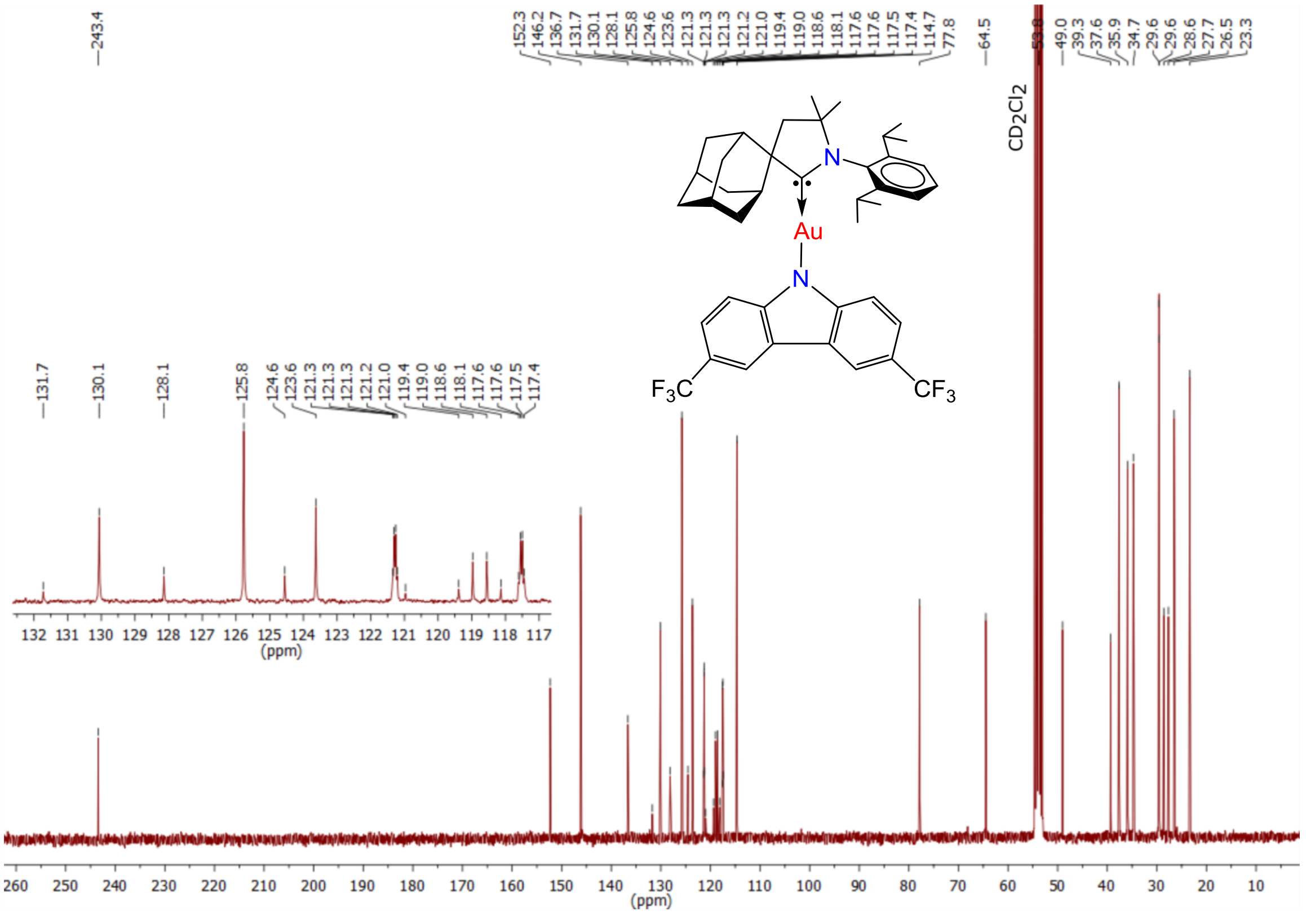

$^{13}C$ NMR (75 MHz, $CD_2Cl_2$)

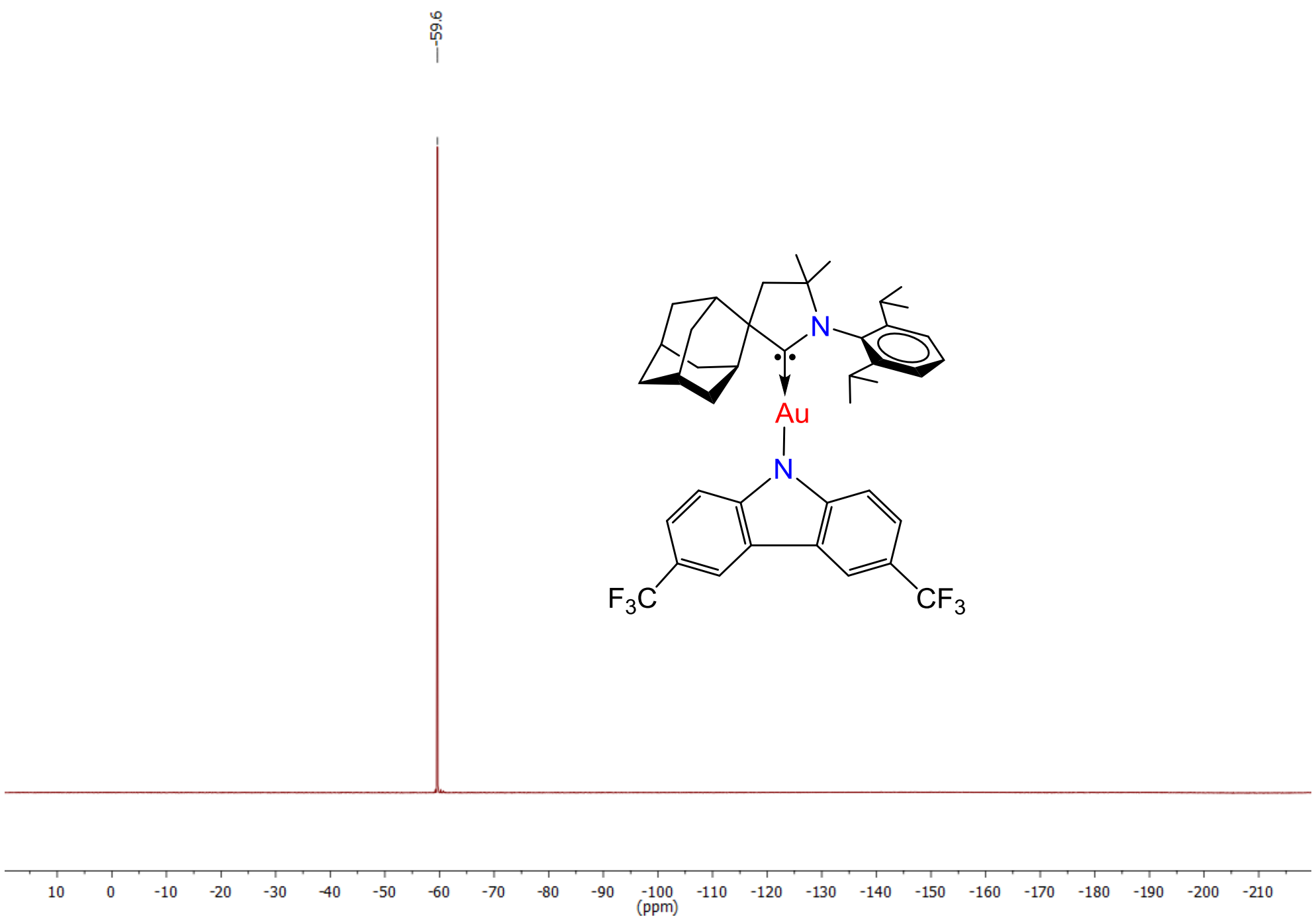

$^{19}$F NMR (282 MHz, $CD_2Cl_2$)

**X-Ray Crystallography.**

Crystals of **1** and **2** suitable for X-ray diffraction study were obtained by layering a toluene solution with hexanes. Complex **1** and **2** crystallizes with two independent molecules in the unit cell and molecules of toluene. The $CF_3$-group was disordered over two half-populated positions for the complex **1** (independent molecule A). For the final refinement, the contribution of severely disordered toluene molecules in the crystals of **1** was removed from the diffraction data with PLATON/SQUEEZE.[7,8] Crystals were mounted in oil on glass fiber and fixed on the diffractometer in a cold nitrogen stream. Data were collected using an Oxford Diffraction Xcalibur-3/Sapphire3-CCD diffractometer with graphite monochromated Mo $K_\alpha$ radiation ($\lambda$ = 0.71073 Å) at 140 K. Data were processed using the CrystAlisPro-CCD and –RED software.[9] The structure was solved by direct methods and refined by the full-matrix least-squares against $F^2$ in an anisotropic (for non-hydrogen atoms) approximation. All hydrogen atom positions were refined in isotropic approximation in a “riding” model with the $U_{iso}$(H) parameters equal to 1.2 $U_{eq}(C_i)$, for methyl groups equal to 1.5 $U_{eq}(C_{ii})$, where $U(C_i)$ and $U(C_{ii})$ are respectively the equivalent thermal parameters of the carbon atoms to which the corresponding H atoms are bonded. All calculations were performed using the SHELXTL software.[10]

The principal crystallographic data and refinement parameters:

Complex **1**, CCDC number 1916996, $C_{51}H_{62}AuF_3N_2$, Monoclinic, space group *P2₁/n*, $a$ = 23.0603(9) Å, $b$ = 15.7169(5) Å, $c$ = 26.0909(9) Å, β = 103.114(4)°, $V$ = 9209.7(6) Å$^3$, $Z$ = 8, $d_{calc}$ = 1.380 g cm$^{-3}$, $\mu$ = 3.242 mm$^{-1}$, colorless/block, crystal size 0.25 × 0.21 × 0.09 mm, $F$(000) = 3904, $T_{min}/T_{max}$ 0.73420/1.00000, $R_1$ = 0.0341 (from 18085 unique reflections with $I$>2σ($I$); $R_{int}$ = 0.0449, $R_{sigma}$ = 0.0392) and $wR_2$ = 0.0852 (from all 73935 unique reflections), *GOF* = 1.096, $\Delta\rho_{min}/\Delta\rho_{max}$ = 1.65/–1.28.

Complex **2**, CCDC number 1916995, $C_{41}H_{45}AuF_6N_2$, Triclinic, space group *P-1*, $a$ = 13.2995(6) Å, $b$ = 14.8662(6) Å, $c$ = 26.4244(10) Å, α = 74.133(4)°, β = 77.531(3)°, γ = 71.883(4)°, $V$ = 4727.0(4) Å$^3$, $Z$ = 4, $d_{calc}$ = 1.232 g cm$^{-3}$, $\mu$ = 3.160 mm$^{-1}$, colorless/block, crystal size 0.16 × 0.11 × 0.07 mm, $F$(000) = 1752, $T_{min}/T_{max}$ 0.70618/1.00000, $R_1$ = 0.0396 (from 18562 unique reflections with $I$>2σ($I$); $R_{int}$ = 0.0548, $R_{sigma}$ = 0.0852) and $wR_2$ = 0.0754 (from all 41002 unique reflections), *GOF* = 1.022, $\Delta\rho_{min}/\Delta\rho_{max}$ = 1.15/–1.28.

**Photophysical Characterisation**

Solution UV-visible absorption spectra were recorded using a Perkin-Elmer Lambda 35 UV/vis spectrometer. UV-vis spectra of solid films were recorded using an Agilent 8453 UV-visible spectrophotometer, with a deuterium-discharge lamp and tungsten lamp, for a wavelength range of 200-1000 nm. Photoluminescence measurements for MeTHF solutions at 298 and 77K were recorded on a Fluorolog Horiba Jobin Yvon spectrofluorimeter. Additional photoluminescence spectra were recorded using an Edinburgh Instruments FLS980 spectrometer. Photoluminescent quantum yield was measured for toluene solutions using an Edinburgh Instruments FS5 spectrometer with 350 nm excitation wavelength for complex **1** (1 mg mL$^{-1}$) and **2** (0.3 mg mL$^{-1}$), and 400 nm excitation wavelength for complexes **3** (0.5 mg mL$^{-1}$) and **4** (0.5 mg mL$^{-1}$). Toluene solutions have been prepared in a nitrogen glovebox from freshly distilled toluene and measured in a 1 cm screw-cap quartz cuvette.

**Transient PL Measurements**

Time-resolved PL spectra of solid films were recorded using an electrically-gated intensified charge-coupled device (ICCD) camera (Andor iStar DH740 CCI-010) connected to a calibrated grating spectrometer (Andor SR303i). Pulsed 400 nm photoexcitation was provided by second harmonic generation (SHG) in a BBO crystal from the fundamental 800 nm output (pulse width = 80 fs) of a Ti:Sapphire laser system (Spectra Physics Solstice), at a repetition rate of 1 kHz. A 425 nm or 375 nm long-pass filter (Thorlabs) was used to prevent scattered laser signal from entering the camera. Temporal evolution of the PL emission was obtained by stepping the ICCD gate delay with respect to the excitation pulse. The minimum gate width of the ICCD was ~5 ns.

The toluene solution time resolved fluorescence data at 298 K was collected on a time-correlated single photon counting (TCSPC) Fluorolog Horiba Jobin Yvon spectrofluorimeter using Horiba Jobin Yvon DataStation v2.4 software. A NanoLED of 370 nm was used as excitation source, with an instrument response function width of 2 ns. The data was analyzed on a Horiba Jobin Yvon DAS6 v6.3 software.

The neat film and host-guest time resolved fluorescence data at 77K were collected on an Edinburgh Instruments FS5 spectrofluorimeter using the 5 W microsecond Xe flash lamp with a repetition rate of 100 Hz (360 nm excitation wavelength).

**OLED fabrication**

OLED devices were fabricated by high-vacuum ($10^{-7}$ Torr) thermal evaporation on ITO-coated glass substrates with sheet resistance of 15 Ω/□. For Architecture A a 40 nm layer of 1,1-bis[4-[N,N-di(4-tolyl)amino]phenyl]-cyclohexane (TAPC) was used, with a 5 nm layer of 9,9'-biphenyl-2,2'-diylbis-9H-carbazole (o-CBP) acting as an exciton blocking layer. The 30 nm thick emissive layer (EML) was composed of either pure **1** and **2** in a host-free configuration or with the emitting material doped at 20 vol.% in either a bis[2-diphenylphosphino)-phenyl]ether oxide (DPEPO) or o-CBP host. A 40 nm layer of diphenyl-4-triphenylsilyl-phenylphosphine oxide (TSPO1) was used as electron-transport and hole-blocking layer. Architecture B: Devices containing the emitters **3** or **4** were fabricated in a host-free configuration or doped at 20 wt.% in either 1,3,5-tris(carbazol-9-yl)benzene (TCP) or DPEPO using a 10 nm layer of 1,4-bis(triphenylsilyl)benzene (UGH2) as a hole-blocking layer and 1,3,5-tris(2-N-

phenylbenzimidazole-1-yl)benzene (TPBi) as electron-transport layer. o-CBP was synthesized according to the literature procedure.[11] TAPC, TCP and UGH2 were purchased from Luminescence Technology Corp. TPBi, DPEPO and TSPO1 were purchased from Shine Materials. All purchased materials were used as received. OLED current density-voltage measurements were made using a Keithley 2400 source-meter unit. The luminance was measured on-axis using a 1 $cm^2$ calibrated silicon photodiode at a distance of 15 cm from the front face of the OLED. Electroluminescence spectra were measured using a calibrated OceanOptics Flame spectrometer. These were found to be constant across the current density ranges reported and found to be stable over the time scale of measurements (~1 minute).

**Supplementary Figures**

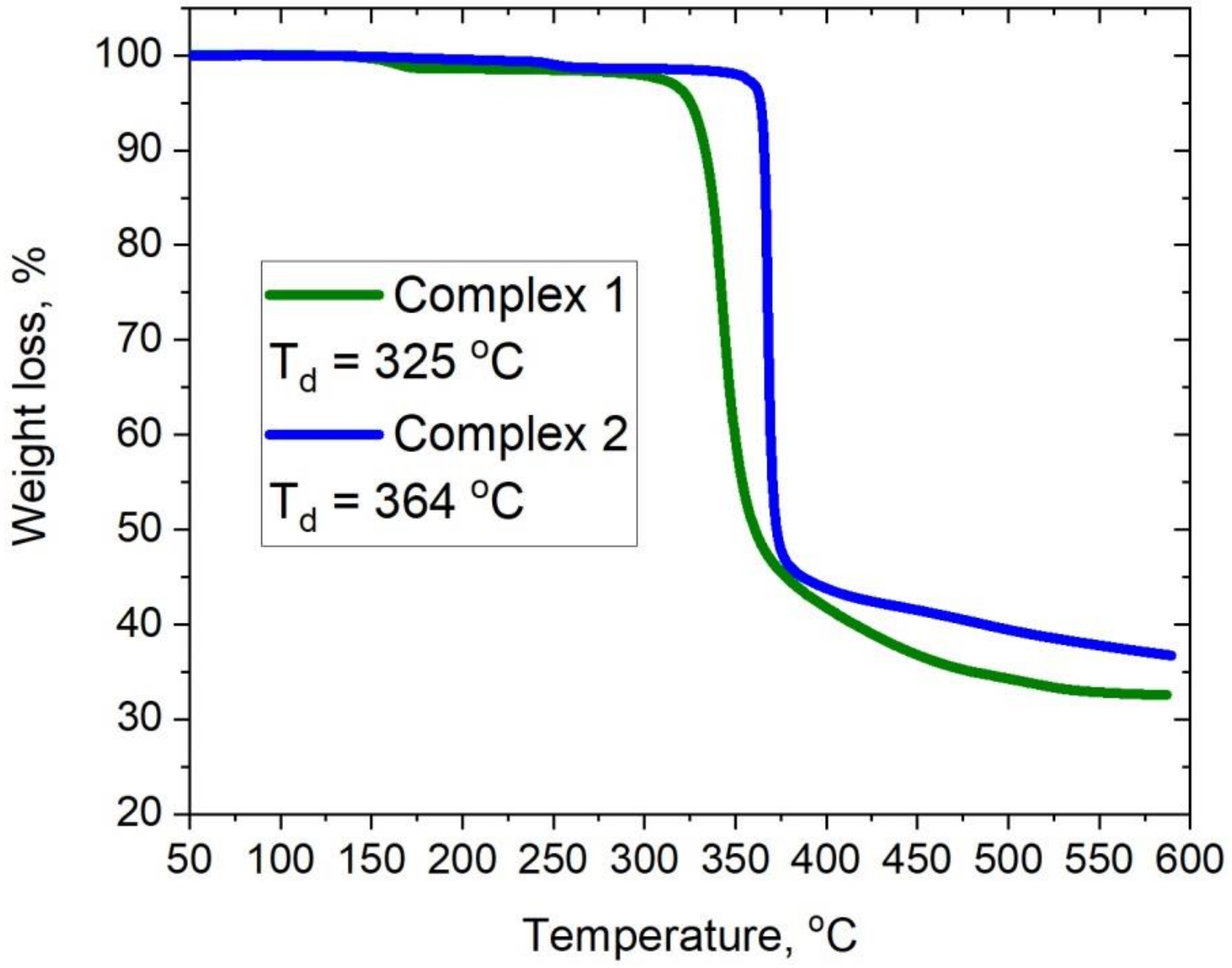

**Figure S1.** TGA curves for gold complexes **1** and **2**.

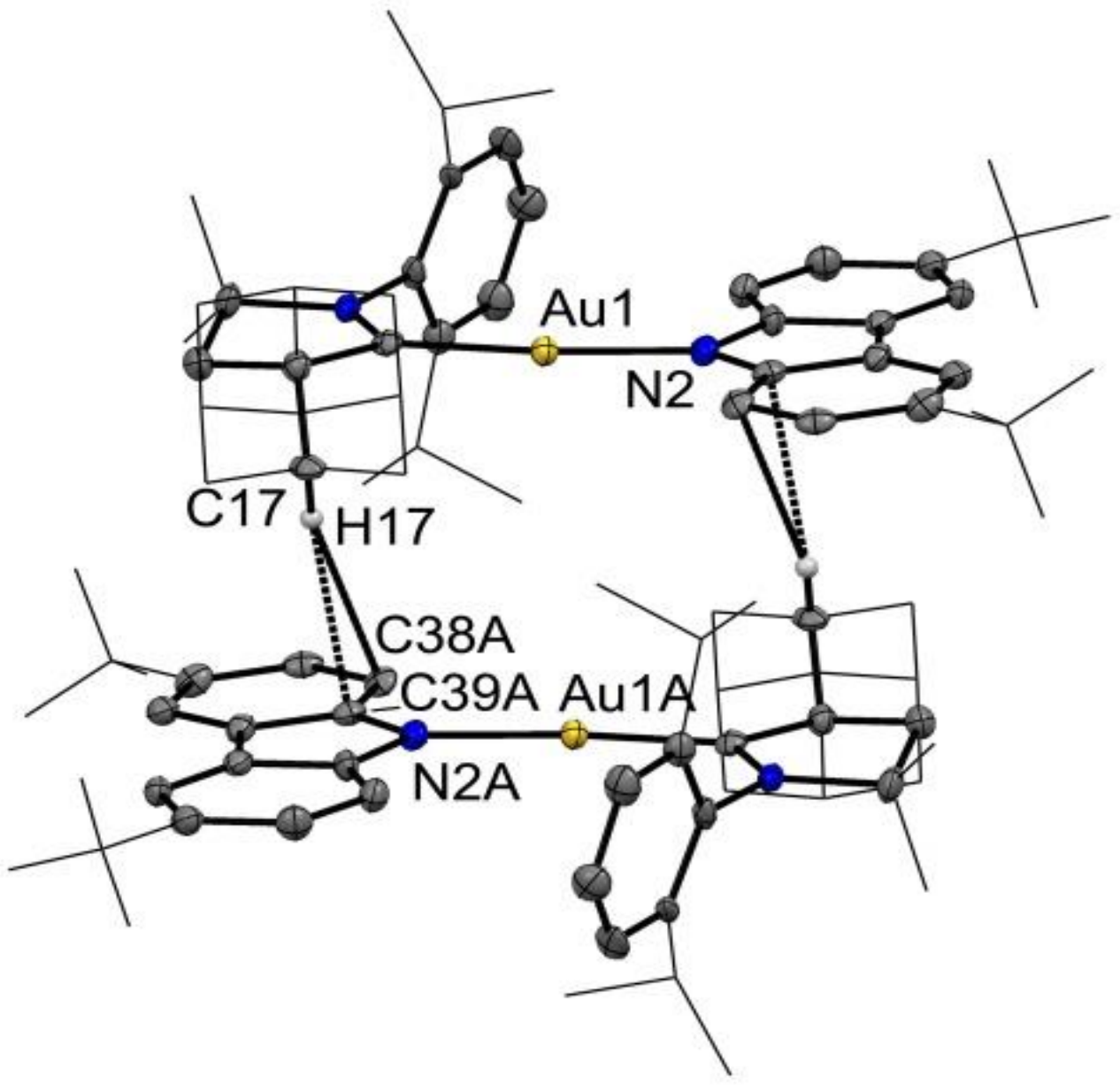

**Figure S2.** The scheme for the intermolecular C–H···π interaction for complex **2**. Ellipsoids are shown at 50 % probability.

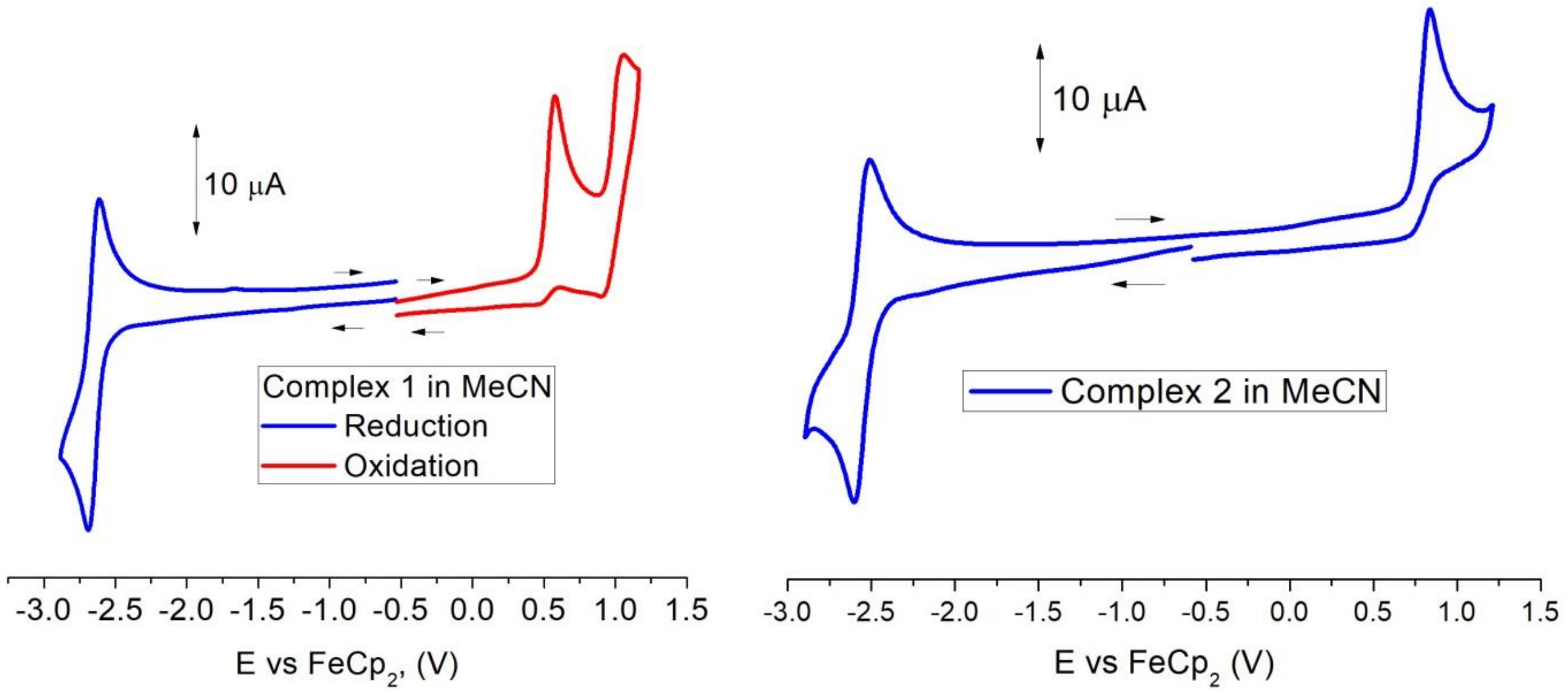

**Figure S3.** Left: reduction (blue) and oxidation (red) cyclic voltammetry scans for gold complex **1**. Right: Full range cyclic voltammogram for gold complex **2**. Recorded using a glassy carbon electrode in MeCN solution (1.4 mM) with $[n\text{-}Bu_4N]PF_6$ as supporting electrolyte (0.13 M), scan rate 0.1 $Vs^{-1}$.

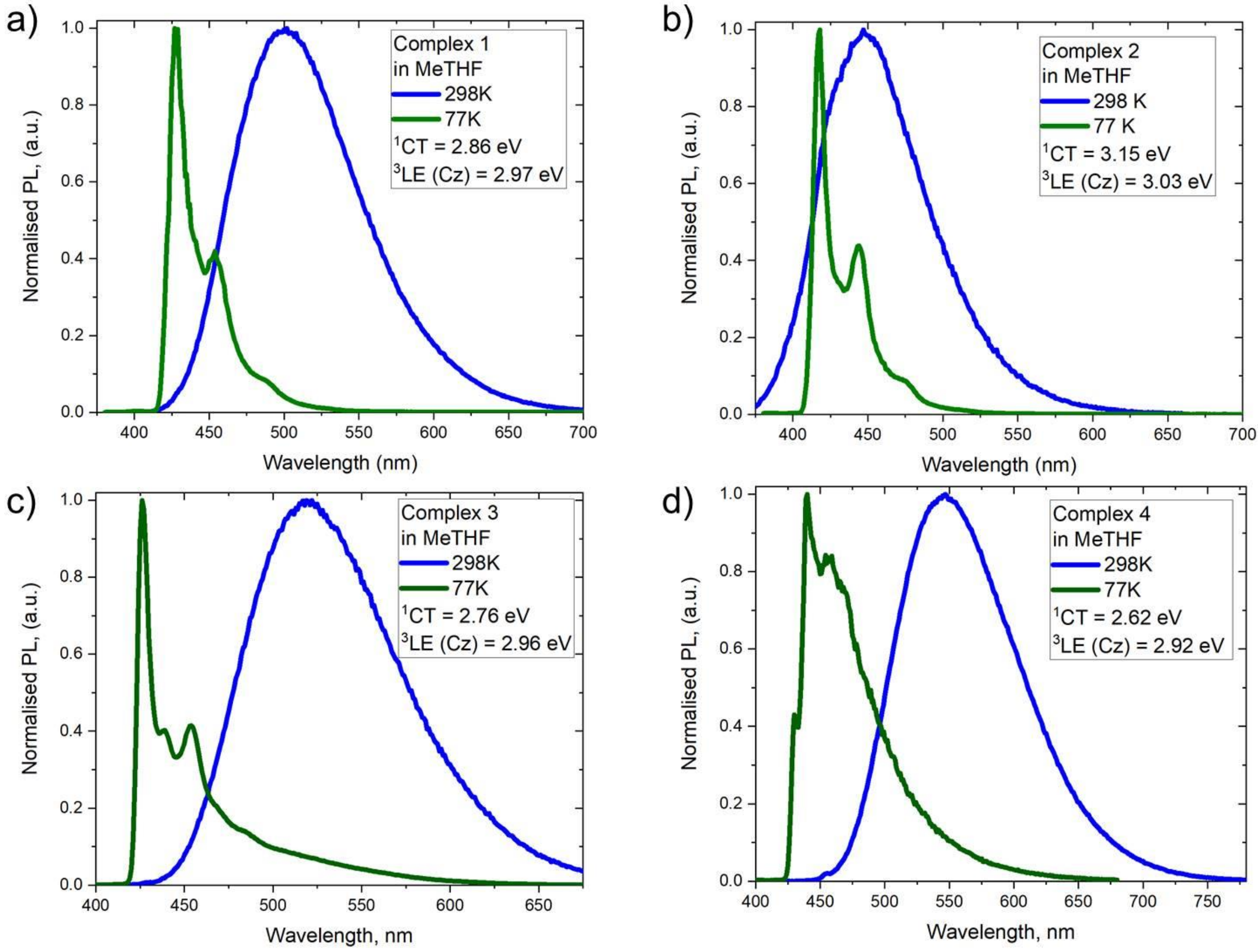

**Figure S4.** Emission spectra for complexes **CMA1** (a), **CMA4** (b), **1** (c) and **2** (d) in MeTHF solutions at 77 and 298K (excitation at 360 nm, under nitrogen).

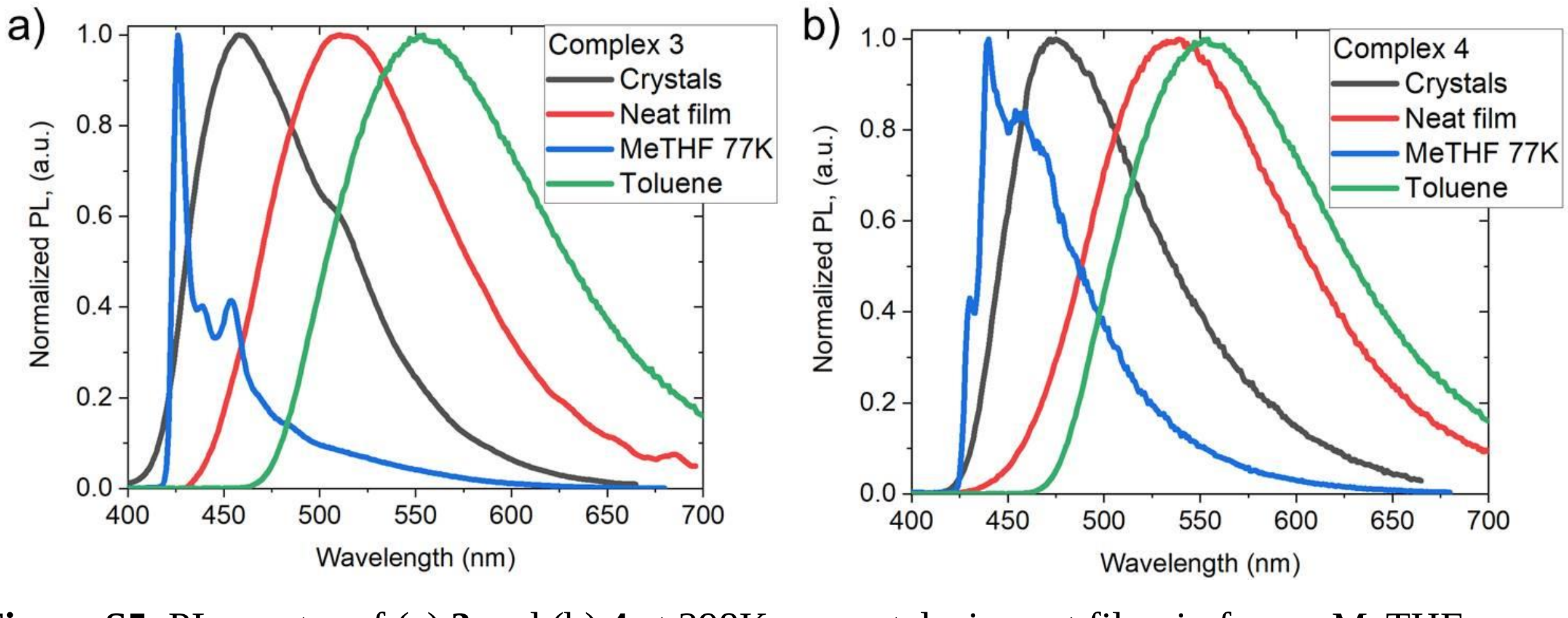

**Figure S5.** PL spectra of (a) **3** and (b) **4** at 298K as crystals, in neat film, in frozen MeTHF solution and in liquid toluene solution (excitation at 365 nm).

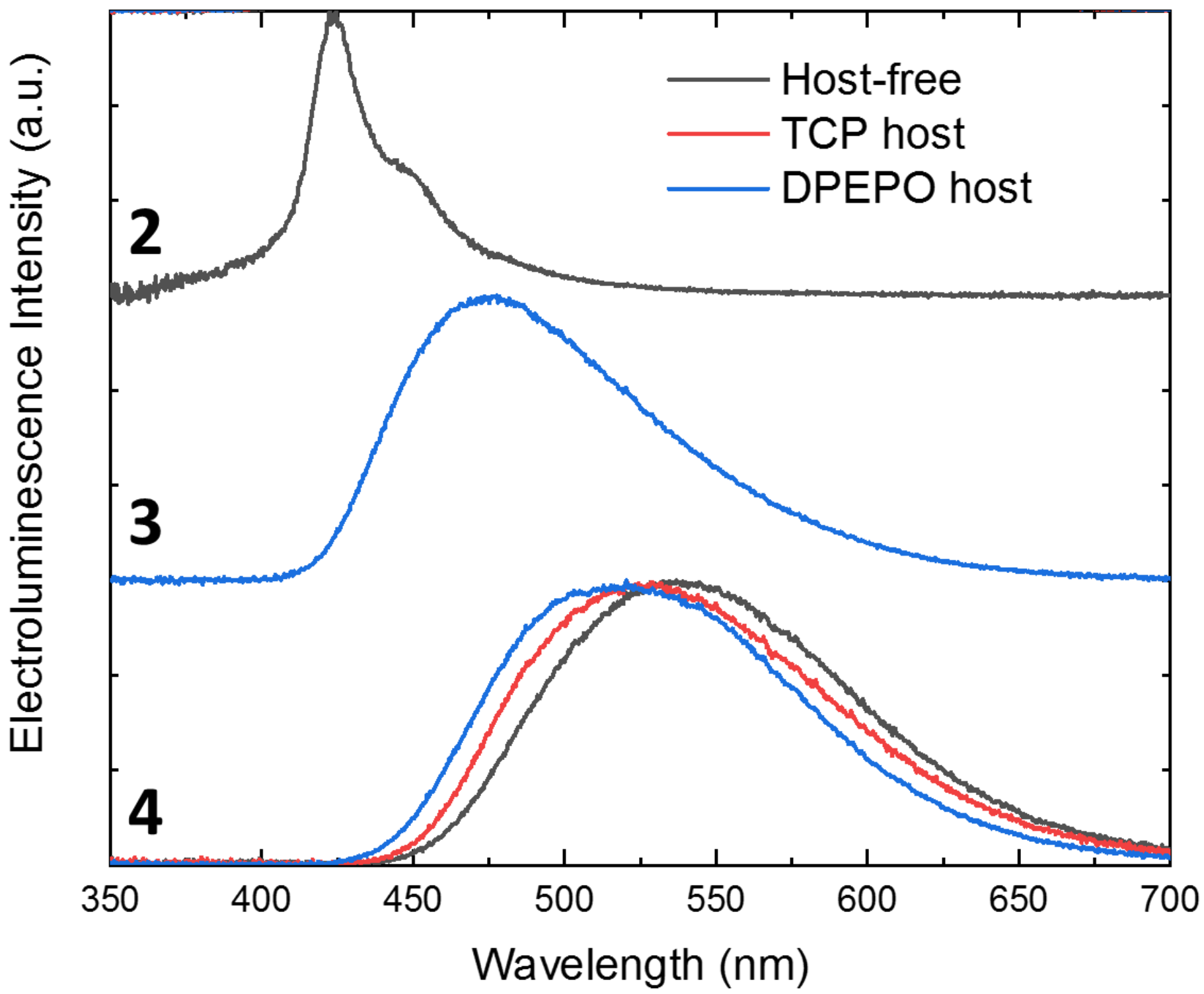

**Figure S6.** Normalised electroluminescence spectra for devices incorporating **2**, **3** and **4** in host-free and host-guest structures.

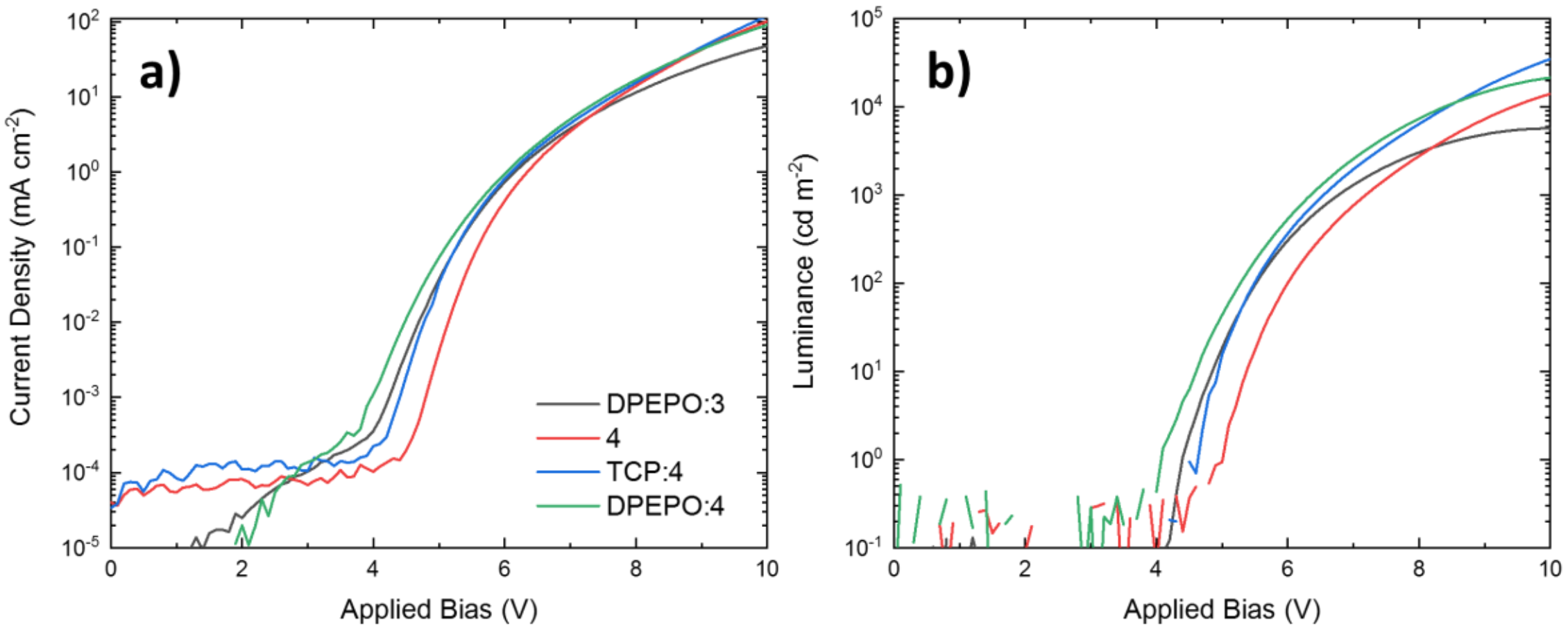

**Figure S7.** a) Current density-voltage and b) luminance-voltage characteristics for OLEDs based on **3** and **4**

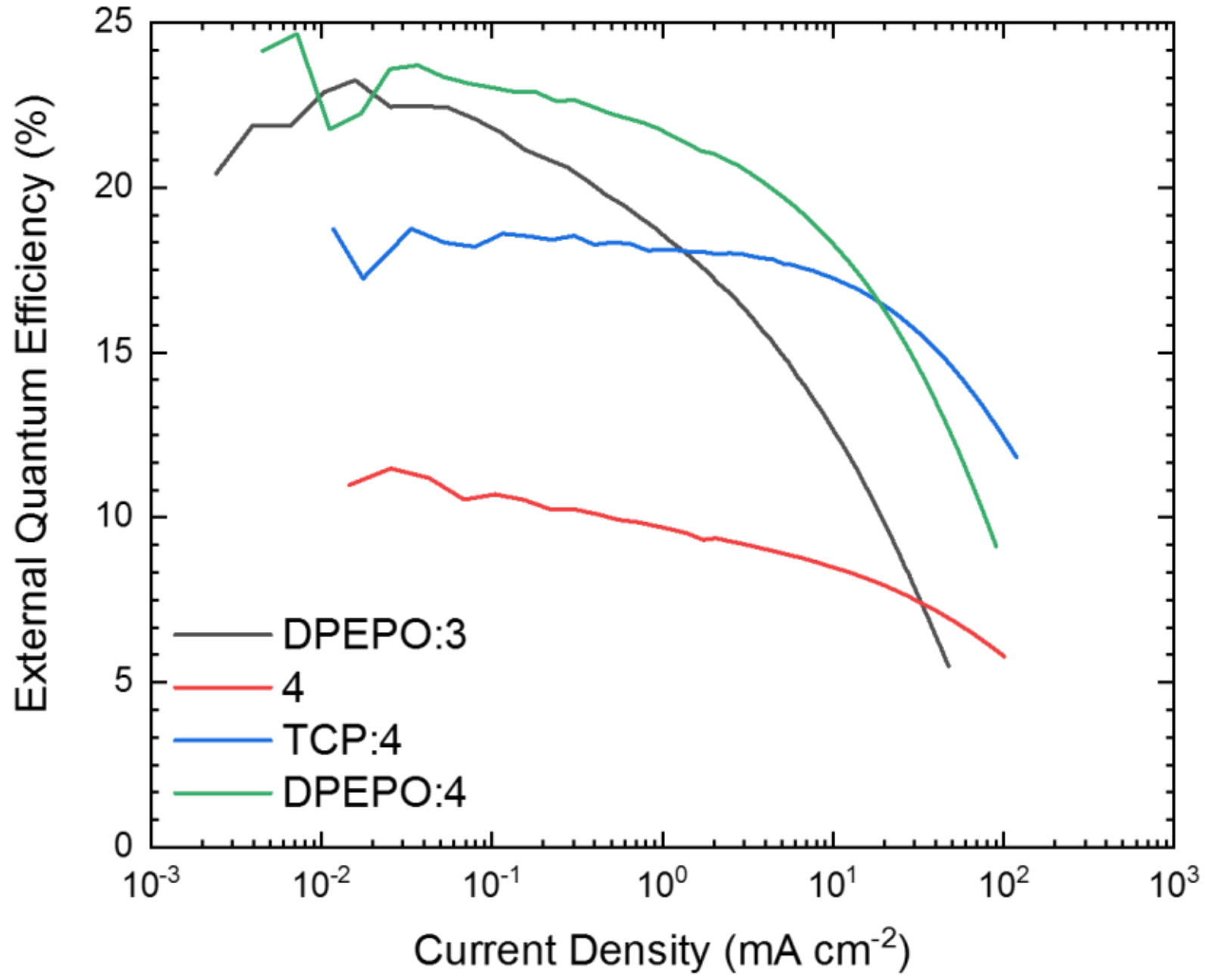

**Figure S8.** External Quantum efficiency as a function of current density for OLEDs based on **3** and **4**

## REFERENCES

([1]) Lavallo, V.; Canac, Y.; Prasang, C.; Donnadieu, B.; Bertrand, G. Stable cyclic (alkyl)(amino)carbenes as rigid or flexible, bulky, electron-rich ligands for transition-metal catalysts: a quaternary carbon atom makes the difference. *Angew. Chem., Int. Ed.* **2005**, *44*, 5705–5709.

(2) Jazzar, R.; Dewhurst, R. D.; Bourg, J.-B.; Donnadieu, B.; Canac, Y.; Bertrand, G. Intramolecular "hydroiminiumation" of alkenes: application to the synthesis of conjugate acids of cyclic alkyl amino carbenes (CAACs). *Angew. Chem., Int. Ed*. **2007**, *46*, 2899–2902.

(3) Jazzar, R.; Bourg, J.-B.; Dewhurst, R. D.; Donnadieu, B.; Bertrand, G. Intramolecular "hydroiminiumation and -amidiniumation" of alkenes: a convenient, flexible, and scalable route to cyclic iminium and imidazolinium salts. *J. Org. Chem*. **2007**, *72*, 3492–3499.

([4]) M. Gantenbein, M. Hellstern, L.L. Pleux, M. Neuburger, M. Mayor, New 4,4′-Bis(9-carbazolyl)−Biphenyl Derivatives with Locked Carbazole−Biphenyl Junctions: High-Triplet State Energy Materials. *Chem. Mater.* **2015**, *27*, 1772−1779.

([5]) Romanov, A. S.; Bochmann, M. Gold(I) and Gold(III) Complexes of Cyclic (Alkyl)(amino)carbenes. *Organometallics* **2015**, *34*, 2439–2454.

([6]) Gritzner, G.; Kůta, J. Recommendations on reporting electrode potentials in nonaqueous solvents: IUPC commission on electrochemistry. *Electrochim. Acta* **1984**, *29*, 869–873.

([7]) Spek, A. L., *Acta Cryst.* **2009**, *D65*, 148.

(8) Sluis, P. van der, Spek, A. L., *Acta Cryst.* **1990**, *A46*, 194.

([9]) *Programs CrysAlisPro*, Oxford Diffraction Ltd., Abingdon, UK (2010).

([10]) Sheldrick, G.M. SHELX-97 and SHELX-2018/3 – Programs for crystal structure determination (SHELXS) and refinement (SHELXL), *Acta Cryst.* **2008**, *A64*, 112; *Acta Cryst.* **2015**, *C71*, 3–8.

(11) Zyryanov, G.V., Kovalev, I.S., Egorov, I.N., Rusinov, V.L., Chupakhin, O.N., Chem. Heterocycl. Comp., 2011, 47(5), 571–574. Translated from Khimiya Geterotsiklicheskikh Soedinenii, No. 5, pp. 692–695, May, 2011.